\documentclass{article}

\usepackage{PRIMEarxiv}

\usepackage[utf8]{inputenc} 
\usepackage[T1]{fontenc}    
\usepackage[hidelinks]{hyperref}       
\usepackage{url}            
\usepackage{booktabs}       
\usepackage{amsfonts}       
\usepackage{nicefrac}       
\usepackage{microtype}      
\usepackage{lipsum}
\usepackage{fancyhdr}       
\usepackage{graphicx}       
\graphicspath{{media/}}     

\usepackage{amsmath,amsthm,amssymb,amsfonts}
\usepackage{float}
\usepackage{tikz}
\usetikzlibrary{calc}
\usetikzlibrary{arrows.meta, patterns, positioning}
\usetikzlibrary{decorations.markings}
\usepackage{subcaption}
\title{Numerical Modeling of Liquid Wall Flows for Fusion Energy Applications Using Maxwell-Navier-Stokes Equations
}

\author{
  Suresh Murugaiyan\\
  Department of Aerospace Engineering \\
  Virginia Tech \\
  Blacksburg, VA 24061 \\
  \texttt{sureshm@vt.edu} \\
  \And
  Stefano Brizzolara  \\
  Department of Aerospace Engineering \\
  Virginia Tech \\
  Blacksburg, VA 24061 \\
  \texttt{stebriz@vt.edu} \\
}

\begin{document}
\maketitle

\begin{abstract}
During the Z-Pinch fusion process, electric current is injected into liquid metal from the plasma column, generating Lorentz forces that deform the liquid metal's free surface. Modeling this phenomenon is essential for assessing the feasibility of using liquid metal as an electrode wall in fusion devices. Traditionally, such problems, where liquid metal is exposed to electromagnetic forces, are modeled using magneto-hydrodynamic (MHD) formulation, which is more suitable for cases without external electric current penetration into liquid metals. MHD formulation typically models situations where liquid metal flows in the presence of an external magnetic field, with the initial magnetic field known and evolving over time via the magnetic induction equation. However, in Z-Pinch fusion devices, the electric current penetrates and traverses through the liquid metal, necessitating numerical calculations for the initial magnetic field. 
Additionally, the deformation of the liquid metal surface alters the current path's geometry and the resulting magnetic field, rendering traditional MHD formulations unsuitable. This work addresses this issue by directly solving Maxwell's equations, instead of the magnetic induction equation, in combination with Navier-Stokes equations, making it possible to predict the magnetic field even when the fluid is in motion. The Maxwell equations are solved in potential formulation alongside Navier-Stokes equations using a finite volume numerical method on a collocated grid arrangement. This proposed numerical framework successfully captures the deformation of the liquid metal's free surface due to the applied electric current.
\end{abstract}



\section{Introduction}
The quest for fusion energy aims to deliver a clean, abundant, and sustainable power source that can meet global energy demands without the environmental drawbacks of fossil fuels. Among the different methods for achieving controlled thermonuclear fusion, magnetic confinement fusion, particularly the Z-Pinch, stands out as a compelling and promising area of research. Z-Pinch involves passing a large current through a plasma column, creating a very strong magnetic field that helps confine and compress the plasma to conditions suitable for fusion.

Within this plasma, Deuterium and Tritium, two hydrogen isotopes, are key players in one of the most common methods to achieve fusion. When these isotopes are compressed and heated to millions of degrees, their nuclei undergo fusion, producing helium and a neutron. This fusion reaction releases a substantial amount of energy (17.6 MeV), which can be captured and harnessed. The reaction is represented by the following equation:

\begin{equation*}
    {}^2_1D + {}^3_1T \rightarrow {}^4_2He (3.5 \text{ MeV}) + {}^1_0n (14.1 \text{ MeV})
\end{equation*}

\clearpage
After the fusion reaction, the plasma emits high-energy neutrons that play a crucial role in the reactor's energy production. When these neutrons strike the blanket surrounding the fusion reactor, they transfer their kinetic energy, significantly heating the blanket. The blanket, as the immediate interface with the high-energy plasma and neutrons, endures constant bombardment, leading to severe stress and radiation damage. This results in wear, degradation, and potential impurity introduction into the plasma. Consequently, the need for frequent maintenance and replacement of the blanket impacts the reactor's efficiency and operational lifespan.

Using a liquid wall in fusion reactors offers a promising alternative to traditional solid walls. Unlike solid materials, which can crack and degrade over time, a liquid wall continuously renews its surface, effectively handling the damage from radiation and heat. The flowing liquid metal absorbs and dissipates the intense heat from plasma and neutron impacts more efficiently than solid materials.

Although using a liquid metal wall in Z-Pinch fusion reactors offers potential advantages, its practical implementation requires a thorough understanding of the liquid metal's free surface behavior. This behavior is affected by Z-Pinch plasma currents and various electromagnetic forces present in the fusion environment. Ensuring stable free surface conditions in the liquid metal walls of Z-Pinch devices is crucial for effective fusion, as any instability can lead to plasma contamination and disrupt the fusion process. This paper aims to develop a mathematical model that describes how Z-Pinch plasma currents influence the free surface behavior of the liquid wall.

Various studies have explored the dynamics of fusion liquid walls under these forces across different reactor designs. In a key study, Brannick \cite{brannick2022theory} formulated an analytical model to describe the wave generation in liquid metal induced by the magnetic pressure from a plasma column. In Z-Pinch fusion, electrical currents flow from the plasma column into the surrounding liquid metal. This current flow through the liquid metal produces a spatially varying magnetic pressure, which in turn generates a Lorentz force. Brannick's analytical model examines how this Lorentz force displaces the free surface of the liquid metal electrode, leading to wave formation.

In the study conducted by Sun et al. \cite{sun2023magnetohydrodynamics}, the effects of magneto-hydrodynamic (MHD) drag on flowing liquid metal plasma-facing components were investigated. Although liquid metal is a promising technology for fusion reactors, MHD drag reduces the flow speed in conducting liquids. Experiments were conducted using an open-channel device with either insulating or conducting walls. The results indicated that the average channel flow speed decreased with the use of conducting walls and with increasing transverse magnetic field strength.

Liquid metal plasma-facing components (LM-PFCs) often rely on the Lorentz force to ensure liquid metal adheres to the interior surfaces of fusion reactors. In a study by Hvasta et al. \cite{hvasta2017demonstrating}, the potential of electromagnetic control for managing free-surface liquid metal flows was examined. The researchers introduced electrical currents into the liquid metal as it traversed a uniform magnetic field. This method effectively controlled both the flow velocity and depth, consistent with theoretical predictions, thereby demonstrating the capabilities of electromagnetic control.

A major concern with implementing liquid metal plasma-facing components in fusion reactors is the risk of liquid metal being ejected into the plasma column during the fusion process \cite{fiflis2016free}. Shallow water theory is applied to simulate wave propagation and evaluate the stability of liquid metal under plasma exposure. This theory indicates that ejections are more likely to happen at the edges, where waves collide with the walls, rather than from the center of the trenches.


Many studies analyzing the stability of liquid metal surfaces in fusion plasma have neglected the presence of the plasma sheath region, characterized by large electric fields and strong ion flows. Holgate et al. \cite{holgate2018electrohydrodynamic} address this gap by performing a linear perturbation analysis of the liquid-sheath interface to determine the stability conditions for liquid metal surfaces. Narula et al. \cite{narula2005study} explored the behavior of liquid metal film flow under magnetic fields relevant to fusion environments through a 3D numerical simulation of two-phase electrically conducting liquids. The simulations predicted an increase in liquid metal film thickness due to magneto-hydrodynamic body forces and captured the tendency of the liquid metal to move away from the side walls.

Xiujie et al. \cite{xiujie2008mhd} studied magneto-hydrodynamic effects on liquid metal film flows for plasma-facing components in fusion technology.  They identified three MHD phenomena—retardant, rivulet, and flat film flow—concluding that flat film flow is the most suitable for plasma-facing components due to its resistance to disruptive MHD effects.

Somboonkittichai et al. \cite{somboonkittichai2023surface} formulated a mathematical model employing the Lagrange equation to characterize the deformation of a viscous liquid surface in a magnetized fusion plasma, utilizing Rayleigh’s method. The study concludes that strong magnetic fields, crucial for confinement in fusion devices, primarily govern instability. It suggests reevaluating the use of static planar free liquid surfaces in fusion devices due to their natural instability under these conditions.

Beyond research on fusion liquid walls with free surfaces \cite{murugaiyan2024modeling,murugaiyan2019response}, studies on two-phase magneto-hydrodynamics (MHD) provide significant insights. The investigation of two-phase MHD flows is complex, and this area of research is notably less developed compared to studies on single-phase MHD phenomena.

\clearpage

A numerical framework for multiphase magneto-hydrodynamic flows with large property gradients is studied by \cite{flint2021magneto}. The framework captures Melting, solidification suitable for advanced manufacturing scenarios. Righolt et al. conducted a comprehensive study exploring the analytical solutions for one-way coupled magneto-hydrodynamic free surface flow \cite{righolt2016analytical}. The research examined the behavior of a conductive liquid layer influenced by surface tension, gravity, and Lorentz forces. Kharicha et al. \cite{kharicha2015experimental} studied the deformation of a liquid metal's free surface under electric currents using experimental and numerical methods. A cylindrical container filled with liquid metal had electric currents applied through central copper electrodes. At low current densities, the interface shifts downward due to the Lorentz force's pinch effect. At high current densities, the increased flow causes further interface displacement driven by a Bernoulli mechanism.

Research into the stability of fusion liquid walls has extensively investigated the complex electromagnetic forces present in fusion environments. Many studies have utilized traditional magneto-hydrodynamic (MHD) equations, particularly focusing on the magnetic induction equation to model the magnetic field's evolution. However, Z-pinch scenarios pose unique challenges when plasma current directly interacts with liquid metal. The MHD approach requires the magnetic field generated by the current as an initial condition, which is often unavailable. This issue stems not only from the lack of analytical solutions even for simpler geometries but also from the deformable nature of liquid metals. To overcome these challenges, this research develops a numerical solver that directly solves Maxwell's equations to calculate the magnetic field induced by the current.

The developed numerical solver models the behavior of a quiescent liquid metal layer subjected to an external electric current by directly solving Maxwell's equations alongside the Navier-Stokes equations. This acts as a model for plasma current interacting with liquid metal walls in a Z-pinch fusion device. A simplified computational model is used, where a wire carrying an electric current replaces the plasma column. The current flows through the wire, enters the liquid metal, and then exits through the wall of the cylindrical container, which is grounded.

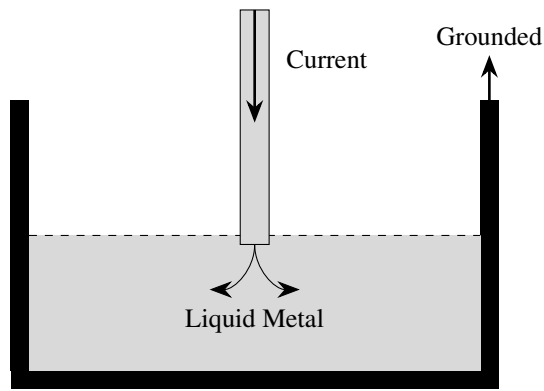
\begin{figure}[H]
	\centering
	\begin{tikzpicture}[scale=1.2, >=Stealth]
		\def\wallheight{3}
		\def\wallthickness{0.2}
		\def\fluidheight{1.5}
		\def\wirewidth{0.32}
		\def\wireheight{2.6} 
		
		\fill[black] (-\wallthickness,0) rectangle (0,\wallheight); 
		\fill[black] (5,0) rectangle (5+\wallthickness,\wallheight); 
		\fill[black] (-\wallthickness,0) rectangle (5+\wallthickness,-\wallthickness); 
		
		\fill[gray!30] (0,0) rectangle (5,\fluidheight); 
		
		\draw[dashed] (0,\fluidheight) -- (5,\fluidheight);
		
		\node at (2.5,\fluidheight/2-0.2) {Liquid Metal};
		
		\draw[fill=gray!30] (2.5-\wirewidth/2,\fluidheight - 0.1) rectangle ++(\wirewidth,\wireheight); 
		
		\draw[-{Stealth[length=3mm]}, line width=1pt, black] (2.5,\fluidheight+\wireheight - 0.1) -- ++(0,-\wireheight/2 + 0.05);
		\node[black] at (3.3,\fluidheight+\wireheight/2+\wireheight/4) {Current}; 
		
		\draw[-{Stealth[length=3mm]}, line width=1pt, black] (5+\wallthickness/2,\wallheight) -- ++(0,0.5) node [above] {Grounded};
		
		\draw[-{Stealth[length=3mm]}, black] (2.5,1.4) to [bend right=45] ++(0.5,-0.5);
		\draw[-{Stealth[length=3mm]}, black] (2.5,1.4) to [bend left=45] ++(-0.5,-0.5);

	\end{tikzpicture}
	\caption{Geometry representing current carrying wire in fusion liquid wall model.}
	\label{fig:1.3}
\end{figure}

The finite volume method, using the PISO algorithm on a collocated grid with Rhie-Chow interpolation, is employed to solve the Navier-Stokes equations. However, Maxwell's equations in their original form include curl operators, which necessitate storing variables at locations other than the cell center—such as on cell edges and vertices—when discretized using finite volume methods. As a result, the original form with curl operators does not fit well within this finite volume framework. Conversely, Maxwell's equations in their potential form, which consist of Laplacian and divergence terms, are more suited to this computational approach.

\section{Mathematical Formulation}
The deformation of a liquid metal surface under applied electric currents is governed by Maxwell's equations and Navier-Stokes equations. Maxwell's equations are presented in their differential form, along with a discussion on the insignificance of displacement current in conductors.

\subsection{Maxwell's Equations} 

In the absence of electric charge, Gauss's Law for electric field is given by:
\begin{equation} 
	\nabla \cdot \textbf{E} = 0
\end{equation}
\clearpage
Faraday's Law of Induction states that a circulating electric field is generated by a magnetic field that is changing over time. 
\begin{equation}  
	\nabla \times \textbf{E} = - \frac{\partial \textbf{B}}{\partial t}
\end{equation}
Gauss's law for Magnetic field can be expressed mathematically as:
\begin{equation} 
	\nabla \cdot \textbf{B} = 0
\end{equation}
The Ampere-Maxwell law can be expressed as:
\begin{equation}
	\nabla \times \mathbf{B} = \mu_0 \mathbf{J} + \mu_0 \epsilon_0 \frac{\partial \mathbf{E}}{\partial t}
\end{equation}

Here $\textbf{E}$ is electric Field $(V/m)$,  $\textbf{B}$ is Magnetic flux density $(T)$, $\textbf{J}$ is conduction current density $(A/m^2)$, $\epsilon_o$ is permittivity of free space and $\mu_o$ is permeability of free space. 

\subsection{Displacement Current in Conductors} 
The displacement current is generally negligible compared to the conduction current in conductors (\cite{lonngren2007fundamentals}).

Consider the Ampere-Maxwell law
\begin{equation*}
	\nabla \times \textbf{B} = \mu_0  \textbf{J}  + \mu_0 \epsilon_o  \frac{\partial \textbf{E}}{\partial t}
\end{equation*}

According to Ohm's law, the conduction current $\mathbf{J}_c$ is expressed in the above equation, $\textbf{J}_c = \textbf{J}  = \sigma \textbf{E}$. Here $\sigma$ is electrical conductivity of the material $(S/m)$.The displacement current in frequency domain is
\begin{equation*}
	\textbf{J}_d = \epsilon_o  \frac{\partial \textbf{E}}{\partial t} = i \omega \epsilon_o \textbf{E} = i 2 \pi f \epsilon_o \textbf{E}
\end{equation*}

Here $\omega$ is angular frequency $(rad/s)$ and $f$ is frequency $(Hz)$.  The magnitude of the conduction current compared to the displacement current is given by the ratio

\begin{equation*}
	\frac{J_c}{J_d} =  \frac{\sigma }{ 2 \pi f \epsilon_o }
\end{equation*}

The conduction current becomes comparable to the displacement current at a certain frequency and is given by the ratio
\begin{equation*}
	f = \frac{\sigma}{2 \pi \epsilon_o}
\end{equation*}
For copper, this frequency is calculated as follows:
\begin{equation*}
	f = \frac{5.8 \times 10^7}{2 \times \pi \times 8.854 \times 10^{-12}} = 1.0425 \times 10^{18} Hz
\end{equation*}

In this work, the displacement current is neglected, making the derived governing equations applicable for frequencies well below $1.0 \times 10^{18}$ Hz. It's important to note that $1.0 \times 10^{18}$ Hz is an extremely high frequency, far beyond the range of most practical applications. This assumption simplifies the Ampere-Maxwell law for conductors, allowing it to be expressed solely in terms of conduction current density.

The simplified form of the Ampere-Maxwell law is:
\begin{equation*} 
	\nabla \times \textbf{B} = \mu_0 \textbf{J}
\end{equation*}

\subsection{Potential Formulation of Maxwell's Equations} 
The potential formulation of Maxwell's equations provides an alternative representation by expressing electromagnetic fields through scalar and vector potentials rather than electric and magnetic fields.  When applying a finite volume method, directly discretizing Maxwell's equations with curl operators typically requires placing variables on cell edges or vertices. In contrast, the current finite volume framework uses a collocated grid with Rhie-Chow interpolation for the Navier-Stokes equations, where variables are stored at cell centers.

\clearpage

Recasting Maxwell's equations in terms of potentials removes the need for curl operators, thus facilitating the use of the collocated grid approach. Detailed discussions of this potential formulation can be found in standard textbooks (\cite{griffiths2017introduction, jackson1999classical}). This formulation is achieved by defining the electric scalar potential $\phi \, (V)$ and the magnetic vector potential $\mathbf{A} \, (V s/m)$.

The electric and magnetic field in terms of these potentials are given by:

\begin{equation}
	\textbf{B} = \nabla \times \textbf{A}
\end{equation}

\begin{equation}
	\textbf{E}   = -\nabla \Phi - \frac{\partial \textbf{A} }{\partial t}
\end{equation}

The definitions of the electric scalar potential ($\phi$) and the magnetic vector potential ($\mathbf{A}$) inherently satisfy Gauss's law for magnetic fields and Faraday's law (\cite{griffiths2017introduction, jackson1999classical}). Consequently, employing these potentials automatically fulfills two of Maxwell's four equations.

By applying these definitions to Maxwell's equations, along with Ohm's law ($\textbf{J} = \sigma \textbf{E}$), the evolution equations for the electric scalar potential and the magnetic vector potential are derived as follows (\cite{griffiths2017introduction, jackson1999classical, murugaiyan2024modeling}):

\begin{equation} \label{eqn:7}
    \nabla \cdot \sigma \nabla \Phi +  \nabla \cdot\left( \sigma \frac{\partial   \textbf{A} }{\partial t} \right)= 0
\end{equation}	
 
 \begin{equation}  \label{eqn:8}
 \mu_0  \sigma \frac{\partial \textbf{A} }{\partial t} - \nabla \cdot \nabla \textbf{A} + \mu_0  \sigma \nabla \Phi = 0
\end{equation}

The equation governing the evolution of the electric potential (equation \ref{eqn:7}) satisfies Gauss's law for electric fields, whereas the equation for the magnetic potential (equation \ref{eqn:8}) satisfies Ampere's law. A comprehensive derivation of these equations is detailed in (\cite{murugaiyan2024modeling}).

\subsection{Potential Formulation of Maxwell Equations For Moving Conductors} 
 For moving conductors such as liquid metal, the Ohm's law is given by

 \begin{equation} 
	\textbf{J} = \sigma (\textbf{E} + \textbf{U} \times \textbf{B})
\end{equation}

The resulting Maxwell's equation in potential formulation is given by:

\begin{equation}
\nabla \cdot \sigma \nabla \Phi + \nabla \cdot \left(\sigma \frac{\partial \textbf{A} }{\partial t}\right) - \nabla \cdot \left( \sigma \textbf{U} \times \textbf{B}\right) = 0
\end{equation}

\begin{equation}
     \mu_0  \sigma \frac{\partial \textbf{A} }{\partial t} - \nabla \cdot \nabla \textbf{A} + \mu_0  \sigma \nabla \Phi  -\mu_0  \sigma  \textbf{U} \times \textbf{B} = 0 
\end{equation}

\subsection{Navier-Stokes Equations for Conducting Fluids}
Electrically conducting fluids are influenced by the Lorentz force when electromagnetic fields are present. Consequently, the fluid momentum equation must be revised to include this extra body force (\cite{chen1984introduction, muller2001magnetofluiddynamics}).  Considering these electromagnetic forces, the modified fluid momentum equation is expressed as:

\begin{equation*}
	\frac{\partial \rho \textbf{U}}{\partial t} +
	\nabla \cdot \left(\rho \textbf{U} \textbf{U} \right) -
	\mu \nabla \cdot \left(\nabla \textbf{U} \right) +
	\nabla p  -
	\mathbf{J}\times \mathbf{B} = 0
\end{equation*}
The continuity equation is given by:
\begin{equation*}
\nabla \cdot \mathbf{U} = 0
\end{equation*}

Here, $\textbf{U}$ is fluid velocity $(m/s)$, $\rho$ is fluid density $(kg/m^3)$, $p$ is pressure ($Pa$) and $\mu$ is fluid viscosity $ (N s/ m^2)$. 

\clearpage

\section{Computational Domain: Wedge Formulation for Axi-Symmetric Problems }
All the numerical simulations presented in this paper are axisymmetric simulations performed using wedge mesh. This approach is particularly well-suited for problems where field do not change in the circumferential direction \cite{weller1998tensorial}. 

\begin{figure}[H]
    \centering
    \begin{subfigure}[h]{0.49\textwidth}
        \centering
        \begin{tikzpicture}[scale=1.0]
		\tikzset{line/.style={draw=blue, thick},
				fillstyle/.style={fill=black, fill opacity=0.2},
				dashedline/.style={draw=blue, thick, dashed}}
			\draw[line] (0,0) ellipse (2cm and 1cm);
			
			\draw[line] (0,0) ellipse (0.5cm and 0.25cm);
			
			\draw[red, dashed] (0,0) -- (0,-4);
			
			\coordinate (InnerTangentPoint) at (0.5,-0.125);
			\coordinate (OuterTangentPoint) at (1.85,-0.375);
			\coordinate (TopInner) at ($(InnerTangentPoint)+(0,0.25)$);
			\coordinate (TopOuter) at ($(OuterTangentPoint)+(0,0.75)$);
			\coordinate (BottomInner) at (InnerTangentPoint);
			\coordinate (BottomOuter) at (OuterTangentPoint);
			
			\fill[fillstyle] (InnerTangentPoint) -- (OuterTangentPoint) -- (TopOuter) -- (TopInner) -- cycle;
			
			\draw[line] (InnerTangentPoint) -- (TopInner);
			\draw[line] (OuterTangentPoint) -- (TopOuter);
			\draw[line] (TopInner) -- (TopOuter);
			\draw[line] (BottomInner) -- (BottomOuter);
			
			\coordinate (LeftVerticalTangent) at (-2, 0); 
			\draw[line] (LeftVerticalTangent) -- ++(0, -4); 
			\coordinate (RightVerticalTangent) at (2, 0); 
			\draw[line] (RightVerticalTangent) -- ++(0, -4); 
			
			\draw[line] (-2,-4) arc (180:360:2cm and 1cm); 
			\draw[dashedline] (-2,-4) arc (180:0:2cm and 1cm); 
        \end{tikzpicture}
        \caption{Real Physical Domain}
    \end{subfigure}
    \begin{subfigure}[h]{0.22\textwidth}
        \centering
        \includegraphics[width=\textwidth,trim={12cm 0.15cm 7.5cm 0.10cm},clip]{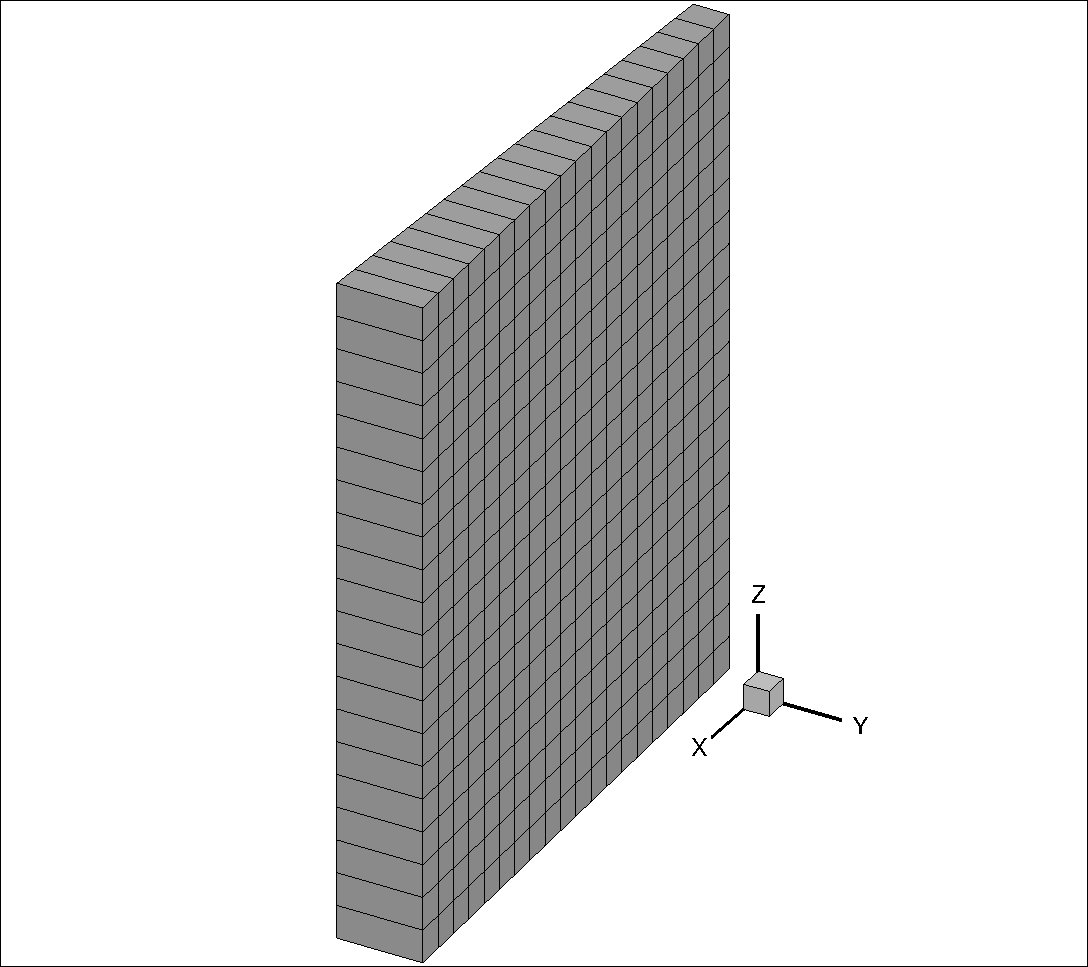}
        \caption{Computational Domain}
    \end{subfigure}
    \hspace{0.4cm}
        \begin{subfigure}[h]{0.24\textwidth}
        \centering
        \includegraphics[width=\textwidth,trim={10cm 0.1cm 7cm 0.10cm},clip]{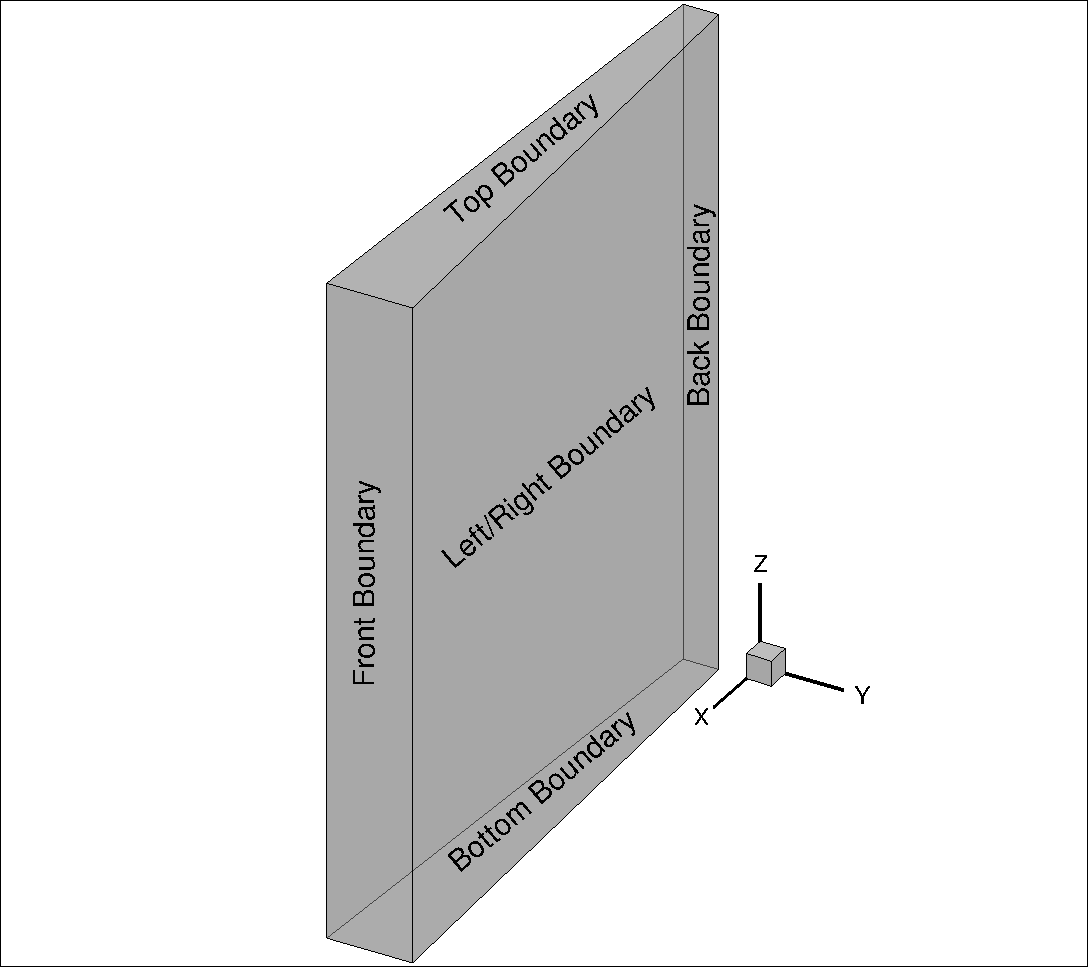}
        \caption{Boundary labels }
            \label{fig:2c}
    \end{subfigure}
    \caption{Computational domain for axially symmetric simulations using wedge mesh \cite{weller1998tensorial}.}
    \label{fig:2}
\end{figure}
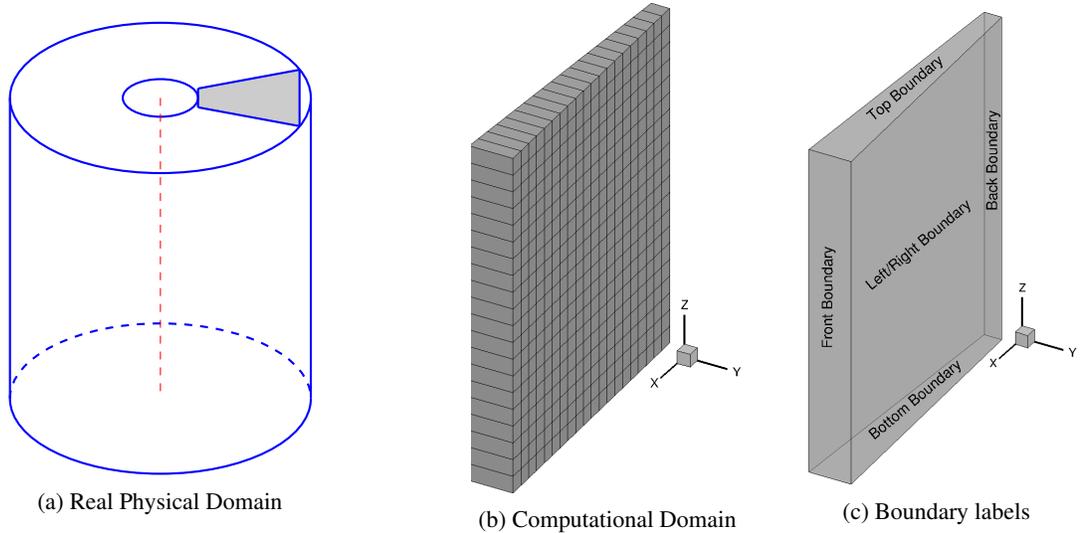

In axially symmetric simulations using wedge meshes \cite{weller1998tensorial} , the computational domain is divided into a series of wedges, each representing a segment of rotation around the central axis. This method facilitates the modeling of the physical system in a 2D plane, thereby reducing the computational load compared to full 3D simulations. A notable advantage of the wedge mesh approach for axially symmetric simulations is that it allows the use of a Cartesian coordinate system for the computations. The wedge method for simulating axially symmetric systems involves modeling the geometry as a thin wedge \cite{weller1998tensorial}. This wedge typically has a small angular span, often less than 5°, and a thickness of a single cell.

In these simulations, the boundaries are defined as illustrated in Figure \ref{fig:2c}. The back boundary represents the inner surface of the cylinder, whereas the front boundary corresponds to its outer surface. The top and bottom boundaries align with the upper and lower surfaces of the cylinder, respectively. Lastly, the left and right boundaries correspond to the axially symmetrical surfaces of the cylinder.

\section{Finite Volume Discretization}
The governing partial differential equations are discretized using the Finite Volume Method \cite{jasak1996error, weller1998tensorial, darwish2016finite}. This method utilizes a grid configuration where all fluid variables are stored at the center of the cells, adhering to a collocated scheme. For more details, refer to Jasak's PhD thesis \cite{jasak1996error}.

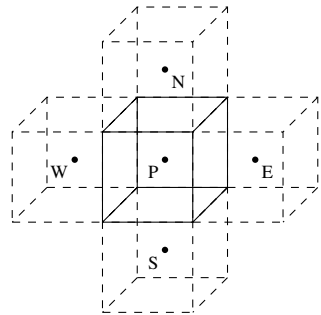
\begin{figure}[H] 
    \centering
    \begin{tikzpicture}[scale=0.6,>=Stealth]
        \scriptsize
        \coordinate (P1) at (0,0,0);
        \coordinate (P2) at (2,0,0);
        \coordinate (P3) at (2,2,0);
        \coordinate (P4) at (0,2,0);
        \coordinate (P5) at (0,0,2);
        \coordinate (P6) at (2,0,2);
        \coordinate (P7) at (2,2,2);
        \coordinate (P8) at (0,2,2);

        \draw (P1) -- (P2) -- (P3) -- (P4) -- cycle; 
        \draw (P5) -- (P6) -- (P7) -- (P8) -- cycle; 
        \draw (P1) -- (P5);
        \draw (P2) -- (P6);
        \draw (P3) -- (P7);
        \draw (P4) -- (P8);

        \coordinate (E1) at (2,0,0);
        \coordinate (E2) at (4,0,0);
        \coordinate (E3) at (4,2,0);
        \coordinate (E4) at (2,2,0);
        \coordinate (E5) at (2,0,2);
        \coordinate (E6) at (4,0,2);
        \coordinate (E7) at (4,2,2);
        \coordinate (E8) at (2,2,2);

        \draw[dashed] (E1) -- (E2) -- (E3) -- (E4) -- cycle; 
        \draw[dashed] (E5) -- (E6) -- (E7) -- (E8) -- cycle; 
        \draw[dashed] (E1) -- (E5);
        \draw[dashed] (E2) -- (E6);
        \draw[dashed] (E3) -- (E7);
        \draw[dashed] (E4) -- (E8);

        \coordinate (W1) at (-2,0,0);
        \coordinate (W2) at (0,0,0);
        \coordinate (W3) at (0,2,0);
        \coordinate (W4) at (-2,2,0);
        \coordinate (W5) at (-2,0,2);
        \coordinate (W6) at (0,0,2);
        \coordinate (W7) at (0,2,2);
        \coordinate (W8) at (-2,2,2);

        \draw[dashed] (W1) -- (W2) -- (W3) -- (W4) -- cycle; 
        \draw[dashed] (W5) -- (W6) -- (W7) -- (W8) -- cycle; 
        \draw[dashed] (W1) -- (W5);
        \draw[dashed] (W2) -- (W6);
        \draw[dashed] (W3) -- (W7);
        \draw[dashed] (W4) -- (W8);

        \coordinate (N1) at (0,2,0);
        \coordinate (N2) at (2,2,0);
        \coordinate (N3) at (2,4,0);
        \coordinate (N4) at (0,4,0);
        \coordinate (N5) at (0,2,2);
        \coordinate (N6) at (2,2,2);
        \coordinate (N7) at (2,4,2);
        \coordinate (N8) at (0,4,2);

        \draw[dashed] (N1) -- (N2) -- (N3) -- (N4) -- cycle; 
        \draw[dashed] (N5) -- (N6) -- (N7) -- (N8) -- cycle; 
        \draw[dashed] (N1) -- (N5);
        \draw[dashed] (N2) -- (N6);
        \draw[dashed] (N3) -- (N7);
        \draw[dashed] (N4) -- (N8);

        \coordinate (S1) at (0,-2,0);
        \coordinate (S2) at (2,-2,0);
        \coordinate (S3) at (2,0,0);
        \coordinate (S4) at (0,0,0);
        \coordinate (S5) at (0,-2,2);
        \coordinate (S6) at (2,-2,2);
        \coordinate (S7) at (2,0,2);
        \coordinate (S8) at (0,0,2);

        \draw[dashed] (S1) -- (S2) -- (S3) -- (S4) -- cycle; 
        \draw[dashed] (S5) -- (S6) -- (S7) -- (S8) -- cycle; 
        \draw[dashed] (S1) -- (S5);
        \draw[dashed] (S2) -- (S6);
        \draw[dashed] (S3) -- (S7);
        \draw[dashed] (S4) -- (S8);

        \fill (1,1,1) circle (2pt); \node[below left] at (1,1,1) {P};
        \fill (3,1,1) circle (2pt); \node[below right] at (3,1,1) {E};
        \fill (-1,1,1) circle (2pt); \node[below left] at (-1,1,1) {W};
        \fill (1,3,1) circle (2pt); \node[below right] at (1,3,1) {N};
        \fill (1,-1,1) circle (2pt); \node[below left] at (1,-1,1) {S};
    \end{tikzpicture}
    \caption{Three-dimensional control volumes in finite volume discretization: Central cell represented by `P' with neighboring cells denoted as `E' (East), `W' (West), `N' (North), and `S' (South).}
\end{figure}

\clearpage

Finite volume discretization uses the integral form of differential operators \cite{haber2014computational, pletcher2012computational, darwish2016finite}, which is derived using Gauss's divergence theorem. This approach transforms the governing differential equations into a matrix equation represented as $AX=B$. The discretization process separates terms into explicit and implicit components, with explicit terms incorporated into vector $B$ and implicit terms included in matrix $A$. Consequently, the algebraic representation for each cell following discretization is formulated as:

\begin{equation}
	a_P \phi_P + \sum_{N=1}^{nfaces} a_N \phi_N = b_P  \label{eq:2.1}
\end{equation}

Where $N$ represents neighboring cells ($E,W,N,S$) and $nfaces$ represents number of cell faces.

\subsection{Integral Definition of Differential Operators}

\subsubsection{Gauss-Divergence Theorem} 
Divergence of vector results in scalar quantity. The integral definition of divergence operator can be written as \cite{jasak1996error,haber2014computational}.
\begin{equation*}
	\nabla \cdot \mathbf{A} = \frac{1}{\Delta V} \int \limits_{V} \nabla \cdot \mathbf{A} \,dV
\end{equation*}
Here, $\mathbf{A}$ is any vector quantity and $\Delta V$ is the volume of the computational cell.
The volume integral can be converted to surface integral using Gauss divergence theorem
\begin{equation*}
\begin{aligned}
\nabla \cdot \mathbf{A} &= \frac{1}{\Delta V}\int \limits_{\partial \Omega} \mathbf{A}\cdot \mathbf{dS}
	&= \frac{1}{\Delta V} \,\,  \sum_{nfaces} \mathbf{A}_f \cdot  \, \textbf{S}
\end{aligned}
\end{equation*}
Here, $ \textbf{S}$ is cell face area vector and $\mathbf{A}_f$ is the face value of the vector $\textbf{A} $.

\subsubsection{Gradient Theorem } 

Gradient of scalar results in vector quantity. The integral definition of gradient operator can be written as
\begin{equation*}
	\nabla \phi = \frac{1}{\Delta V} \int \limits_{V} \nabla \phi \,dV
\end{equation*}
Here, $ \phi$ is any scalar quantity. The volume integral can be converted to surface integral using Gradient theorem
\begin{equation*}
\begin{aligned}
\nabla \phi &= \frac{1}{\Delta V}\int \limits_{ \partial \Omega} \phi \, \mathbf{dS}
	&= \frac{1}{\Delta V} \,\,  \sum_{nfaces} \phi_f \, \, \textbf{S}
\end{aligned}
\end{equation*}
Here, $\phi_f$ is face value of the scalar quantity $\phi$.

\subsection{Discretization of time derivative}
Discretization of time derivative can be carried out by employing either the Euler or the Backward Euler method.

\textbf{Euler Scheme :} The Euler scheme is represented by the following equation ( \cite{weller1998tensorial,jasak1996error} ):
\begin{equation*}
	\frac{\partial \phi }{\partial t}  = \frac{\phi - \phi^o }{\Delta t} 
\end{equation*}
Here, $\phi^o$ represents the value from the previous time step. 
 
 Applying the Euler scheme to each control volume yields the discretized algebraic equation:
\begin{equation*}
	\left( \frac{\partial \phi }{\partial t} \right)_P = \frac{1}{\Delta t} \left(
	 \phi_P -\phi^o_P  \right)
\end{equation*}
The coefficients and source terms for the matrix equation are then calculated as follows:
\begin{equation*}
    \begin{aligned}
        a_P = \frac{1}{\Delta t} &\qquad a_N = 0 &\qquad b_P = \frac{\phi_P ^o}{\Delta t}
    \end{aligned}
\end{equation*}

\clearpage

\textbf{Backward Euler Scheme :} The Backward Euler scheme is expressed by ( \cite{weller1998tensorial,jasak1996error} ):
\begin{equation*}
	\frac{\partial \phi }{\partial t}  = \frac{1}{\Delta t} \left(
	\frac{3}{2} \phi - 2 \phi^o + \frac{1}{2} \phi^{oo} \right)
\end{equation*}
Here, $\phi^{oo} $ represents the value from the second old time step. 

Applying the Backward Euler scheme to each control volume yields the discretized algebraic equation:
\begin{equation*}
	\left( \frac{\partial \phi }{\partial t} \right)_P = \frac{1}{\Delta t} \left(
	\frac{3}{2} \phi_P - 2 \phi^o_P + \frac{1}{2} \phi^{oo}_P \right) = \frac{1}{2 \Delta t} \left(
	3 \phi_P - 4 \phi^o_P +  \phi^{oo}_P \right)
\end{equation*}

The coefficients and source terms for the matrix equation are then calculated as follows:
\begin{equation*}
    \begin{aligned}
        a_P &= \frac{3}{2 \Delta t} &\qquad a_N = 0 &\qquad b_P = \frac{1}{2 \Delta t}  \left( 4  \phi_P ^o -  \phi_P ^{oo} \right)
    \end{aligned}
\end{equation*}

\subsection{Discretization of Laplacian}
The Laplacian term is expressed as:
\begin{equation*}
		\nabla \cdot \sigma \nabla \phi =  \frac{1}{\Delta V} \,\, \int\limits_{V} \nabla \cdot \sigma \nabla \phi \, d V 
\end{equation*}
By applying the divergence theorem, the volume integral is transformed into a surface integral:
\begin{equation*}
	\begin{aligned}
		\nabla \cdot \sigma \nabla \phi
		&= \frac{1}{\Delta V} \,\, \int\limits_{\partial \Omega} (\sigma \nabla \phi )\cdot  \, \textbf{dS} 
		&= \frac{1}{\Delta V} \,\,  \sum_f ( \sigma \nabla \phi )_f \cdot  \, \textbf{S}  \\
	\end{aligned}
\end{equation*}
Gradients must be evaluated at cell faces, which are referred to as surface normal gradients. 

For orthogonal meshes, cell face gradients can be represented as ( \cite{weller1998tensorial,jasak1996error} ):
\begin{equation*}
	\nabla \phi_f \cdot  \, \textbf{S}  =   |\textbf{S}| \frac{\phi_N - \phi_P }{|\textbf{d}|}
\end{equation*}
In this context, $\textbf{d}$ represents the vector denoting the distance from the center of cell P to the center of the neighboring cell N ( \cite{jasak1996error} ). Discretization of Laplacian operator becomes
\begin{equation*}
	\nabla \cdot \sigma \nabla \phi = \frac{1}{\Delta V} \,\,  \sum_f  \sigma_f  |\textbf{S}| \frac{\phi_N - \phi_P }{|\textbf{d}|}   \\
\end{equation*}

Comparing with the matrix equation and gathering the coefficients yields:
\begin{equation*}
	\begin{aligned}
		a_P &= -\frac{1}{\Delta V} \sum_f  \sigma_f   \frac{|\textbf{S}| }{|\textbf{d}|} &\qquad
		a_N &= \frac{1}{\Delta V}  \sum_f  \sigma_f   \frac{|\textbf{S}| }{|\textbf{d}|} &\qquad
		b_P &= 0
	\end{aligned}
\end{equation*}

\paragraph{Dirichlet boundary condition  for Laplacian :} Boundary conditions must be integrated into the algebraic equation system. Dirichlet condition specifies the value of $\Phi$ at the boundary face as $\Phi_b$. For cells with boundary faces:
\begin{equation*}
	\nabla \cdot \sigma \nabla \phi = \frac{1}{\Delta V} \,\,  \sum_{innerfaces}  \sigma_f  |\textbf{S}| \frac{\phi_N - \phi_P }{|\textbf{d}|} +  \frac{1}{\Delta V} \,\,  \sum_{boundaryfaces}  \sigma_b  |\textbf{S}| \frac{\phi_b - \phi_P }{|\textbf{d}|}   \\
\end{equation*}

The matrix coefficients of cells corresponding to boundary faces are:
\begin{equation*}
	\begin{aligned}
		a_P &= -\frac{1}{\Delta V} \,\,  \sum_{boundaryfaces}  \sigma_b   \frac{|\textbf{S}| }{|\textbf{d}|} \qquad
		a_N &= 0 \qquad
		b_P &= - \frac{1}{\Delta V} \,\,   \sum_{boundaryfaces}  \sigma_b   \frac{ |\textbf{S}| }{|\textbf{d}|} \phi_b  
	\end{aligned}
\end{equation*}

\clearpage

\paragraph{Neumann boundary condition for Laplacian :} Neumann boundary condition specifies the value of the gradient of $\Phi$ normal to the boundary face ( \cite{jasak1996error} ):
\begin{equation*}
	\hat{n} \cdot \nabla \Phi = \frac{\textbf{S}}{|\textbf{S}|} \cdot \nabla \Phi = g_b
\end{equation*}
For cells with boundary faces :
\begin{equation*}
	\nabla \cdot \sigma \nabla \phi
	= \frac{1}{\Delta V} \,\,  \sum_{innerfaces} ( \sigma \nabla \phi )_f \cdot  \, \textbf{S} + \frac{1}{\Delta V} \,\,  \sum_{boundaryfaces} ( \sigma \nabla \phi )_b \cdot  \, \textbf{S}  \\
\end{equation*}
For the boundary face:
\begin{equation*}
	\nabla \phi \cdot  \, \textbf{S}  = |\textbf{S}|g_b
\end{equation*}

Then the equation can be written as :
\begin{equation*}
	\nabla \cdot \sigma \nabla \phi
	= \frac{1}{\Delta V} \,\,  \sum_{innerfaces} ( \sigma \nabla \phi )_f \cdot  \, \textbf{S} + \frac{1}{\Delta V} \,\,  \sum_{boundaryfaces} \sigma_b |\textbf{S}|g_b   \\
\end{equation*}
The matrix coefficients of cells corresponding to boundary faces are:
\begin{equation*}
	\begin{aligned}
		a_P &= 0 \quad
		a_N &= 0 \quad
		b_P &= - \frac{1}{\Delta V} \,\,   \sum_{boundaryfaces}  \sigma_b   |\textbf{S}| g_b 
	\end{aligned}
\end{equation*}

\subsection{Discretization of Divergence Term} \label{subsec:4.4}

The Divergence term is expressed as:
\begin{equation*}
	\begin{aligned}
		\nabla \cdot \left(\rho \mathbf{U} \phi\right) &=  \frac{1}{\Delta V} \,\, \int\limits_{V} \nabla \cdot \left(\rho \mathbf{U} \phi\right) \, d V \\
	\end{aligned}
\end{equation*}
By applying the divergence theorem, the volume integral is transformed into a surface integral:
\begin{equation*}
	\begin{aligned}
		\nabla \cdot \left(\rho \mathbf{U} \phi\right)
		&= \frac{1}{\Delta V} \,\, \int\limits_{\partial \Omega} \left(\rho \mathbf{U} \phi\right)\cdot  \, \textbf{dS} 
		= \frac{1}{\Delta V} \,\,  \sum_f ( \rho \mathbf{U} \phi )_f \cdot  \, \textbf{S}  \\
		&= \frac{1}{\Delta V} \,\,  \sum_f \phi _f( \rho \mathbf{U})_f \cdot  \, \textbf{S}  
		= \frac{1}{\Delta V} \,\,  \sum_f \phi _f \,F  
	\end{aligned}
\end{equation*}
The quantity $F = (\rho \mathbf{U})_f \cdot  \, \textbf{S} $ is called mass flux ( \cite{weller1998tensorial,jasak1996error} ). The face value $\phi_f$ is calculated using convection differencing schemes.

For example, the Upwind scheme \cite{weller1998tensorial,jasak1996error} is given by
\begin{equation*}
    \begin{aligned}
    	\phi_f &=  \phi_P \quad \text{F} \geq 0 \\
        \phi_f &=  \phi_N \quad \text{F < 0} \\
	\end{aligned}
\end{equation*}

The two equations can be combined as  \cite{weller1998tensorial} $\phi_f =  \text {max(sign(F),0)}  \, \phi_P + \text {(1-max(sign(F),0))} \, \phi_N$
\begin{equation*}
	\begin{aligned}
		\nabla \cdot \left(\rho \mathbf{U} \phi\right)
		&= \frac{1}{\Delta V} \,\,  \sum_f \phi _f \,F  \\
		&= \frac{1}{\Delta V} \,\,  \sum_f \left(\text {max(sign(F),0)}  \, \phi_P + \text {(1-max(sign(F),0))} \, \phi_N \right)\,F  \\
	\end{aligned}
\end{equation*}
The coefficients are given by:
\begin{equation*}
	\begin{aligned}
		a_P &= \frac{1}{\Delta V} \,\,  \sum_f  \text {max(sign(F),0)} F  \qquad
		a_N &= \frac{1}{\Delta V} \,\,   \sum_f   \text {(1-max(sign(F),0))}  F  \qquad
		b_P &= 0
	\end{aligned}
\end{equation*}

For other higher order convection differencing schemes \cite{weller1998tensorial,jasak1996error} can be referred. 
\clearpage
\subsection{Discretization of Gradient :} 

The gradient term is handled explicitly, which means that it is evaluated based on the previous time step values of the variables.
\begin{equation*}
	\nabla \phi = \frac{1}{\Delta V} \int \limits_{V} \nabla \phi \,dV
\end{equation*}
By applying the divergence theorem, the volume integral is converted to a surface integral:
\begin{equation*}
	\begin{aligned}
		\nabla \phi &= \frac{1}{\Delta V}\int \limits_{ \partial \Omega} \phi \, \mathbf{dS}
		&= \frac{1}{\Delta V} \,\,  \sum_{nfaces} \phi_f \, \, \textbf{S}
	\end{aligned}
\end{equation*}
When comparing with the matrix equation and collecting the coefficients:
\begin{equation*}
\begin{aligned}
	a_P &= 0  \quad a_N &= 0 \quad b_P &= - \frac{1}{\Delta V} \,\,  \sum_{f} \phi_f \, \, \textbf{S}
 	\end{aligned}
\end{equation*}

\section{Model Validation: Test Cases}

\subsection{Steady Axial Current Carrying Wire}
A steady current flows through a long, hollow, straight wire, driven by a constant potential difference between its ends. The wire, characterized by its length \( l \) and inner and outer radii \( r_i \) and \( r_o \), respectively, is modeled as a hollow cylinder with a uniform cross-sectional area. Numerical simulations are employed to calculate various parameters, such as the potential distribution along its length (\( \Phi \)), the surface current density (\( \mathbf{J} \)), the magnetic vector potential (\( \mathbf{A} \)), and the magnetic field (\( \mathbf{B} \)) inside the wire.

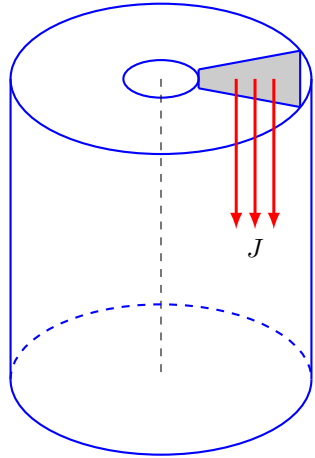
\begin{figure}[H]
	\centering
		\begin{tikzpicture}[scale=1.0]
			
			\tikzset{
				line/.style={draw=blue, thick},
				fillstyle/.style={fill=black, fill opacity=0.2},
				dashedline/.style={draw=blue, thick, dashed}
			}
			
			\draw[line] (0,0) ellipse (2cm and 1cm);
			
			\draw[line] (0,0) ellipse (0.5cm and 0.25cm);
			
			\draw[black, dashed] (0,0) -- (0,-4);
			
			\coordinate (InnerTangentPoint) at (0.5,-0.125);
			\coordinate (OuterTangentPoint) at (1.85,-0.375);
			\coordinate (TopInner) at ($(InnerTangentPoint)+(0,0.25)$);
			\coordinate (TopOuter) at ($(OuterTangentPoint)+(0,0.75)$);
			\coordinate (BottomInner) at (InnerTangentPoint);
			\coordinate (BottomOuter) at (OuterTangentPoint);
			
			\fill[fillstyle] (InnerTangentPoint) -- (OuterTangentPoint) -- (TopOuter) -- (TopInner) -- cycle;
			
			\draw[line] (InnerTangentPoint) -- (TopInner);
			\draw[line] (OuterTangentPoint) -- (TopOuter);
			\draw[line] (TopInner) -- (TopOuter);
			\draw[line] (BottomInner) -- (BottomOuter);
			
			\coordinate (LeftVerticalTangent) at (-2, 0); 
			\draw[line] (LeftVerticalTangent) -- ++(0, -4); 
			\coordinate (RightVerticalTangent) at (2, 0); 
			\draw[line] (RightVerticalTangent) -- ++(0, -4); 
			
			\draw[line] (-2,-4) arc (180:360:2cm and 1cm); 
			\draw[dashedline] (-2,-4) arc (180:0:2cm and 1cm); 
			
			\tikzset{currentarrow/.style={->, >=latex, line width=0.4mm, red}} 

			\newcommand{\currentarrowone}{
			\draw[currentarrow] (1.50,0) -- (1.50,-2.0);
			\node at (1.25,-2.25) {$J$};
			}
			
			\newcommand{\currentarrowtwo}{
			\draw[currentarrow] (1.25,0) -- (1.25,-2.0);
			}
			
			\newcommand{\currentarrowthree}{
			\draw[currentarrow] (1.0,0) -- (1.0,-2.0);
			}
			
			\currentarrowone
			\currentarrowtwo
			\currentarrowthree

		\end{tikzpicture}

	\caption{Hollow cylindrical geometry representing a wire carrying a steady axial current. The current density is uniform across the cross section of the wire. The computational domain is the shaded wedge geometry. Boundary labels and the representative computational mesh is presented in  Figure \ref{fig:2}.}
 \label{fig:3.1}
\end{figure}

The governing equation used for steady fields (equation \ref{eqn:7} and \ref{eqn:8} ) are
\begin{equation} \label{eqn:13}
	\nabla \cdot \sigma \nabla \Phi  = 0
\end{equation}
\begin{equation}  \label{eqn:14}
	\nabla \cdot \nabla \textbf{A} - \mu_0 \sigma \nabla \Phi = 0
\end{equation}

The numerical solutions are obtained by solving non-dimensional form of the governing equations. The non-dimensional form of the governing equations is derived by introducing 
length scale \( L_o \), 
electric conductivity scale be $\sigma_o$,
electric potential scale be $\Phi_o$ and 
magnetic vector potential scale be $A_o$.
Non-dimensional quantities $\nabla\,'$, $\sigma\,'$,  $\Phi\,'$, $\mathbf{A}\,'$  are employed to correspond to $\nabla$, $\sigma$, $\Phi$, $\mathbf{A}$. The dimensional quantities can be recovered by multiplying corresponding scale factors ( For example, $\mathbf{A} = A_o \, \mathbf{A}\,'$ ).

\clearpage

\begin{table}[H]
    \centering
    \caption{Geometry, physical property and boundary condition parameters used for the numerical simulation of  wire carrying a steady current in the axial direction.}
    \begin{tabular}{llll}
        \toprule
        Parameter & Description & Value & Units \\
        \midrule
		$r_i$ & Inner radius of wire & $0.5$ & mm \\
		$r_o$ & Outer radius of wire & $2.5$ & mm \\
		$l$ & Length of wire & $5$ & mm \\
		$\sigma$ & Electrical conductivity & $5.8 \times 10^7$ & S/m \\
		$\mu_0$ & Magnetic permeability & $4 \pi \times 10^{-7}$ & H/m \\
		$\Phi_b$ & Potential at bottom boundary & $0.0$ & V \\
		$\Phi_t$ & Potential at top boundary & $0.5$ & V \\
		$\mathbf{A}_f$ & Vector potential at front boundary & $0.0$ & T $\cdot$ m \\
        \bottomrule
    \end{tabular}
	\label{tab:3.1}
\end{table}

\begin{table}[H]
    \centering
    \caption{Boundary conditions used for the numerical simulation of  wire carrying a steady current in the axial direction.}
    \begin{tabular}{lll}
        \toprule
        & \multicolumn{2}{c}{Boundary Condition}\\
        \cmidrule(r){2-3}
        Boundary & Potential ($\Phi$) & Vector potential ($\mathbf{A}$) \\
        \midrule
		Top boundary & Fixed value & Zero normal gradient \\
		Bottom boundary & Fixed value & Zero normal gradient \\
        Left boundary & Zero normal gradient & Zero normal gradient \\
		Right boundary & Zero normal gradient & Zero normal gradient \\
		Front boundary & Zero normal gradient & Fixed value \\
		Back boundary & Zero normal gradient & Zero normal gradient \\
        \bottomrule
    \end{tabular}
	\label{tab:3.2}
\end{table}

The governing equations in non-dimensional form \cite{murugaiyan2024modeling} is given by

\begin{equation*}
	\nabla'\cdot\sigma'\nabla'\Phi' = 0
\end{equation*}
\begin{equation*}
	\nabla'\cdot\nabla'\textbf{A}' -\sigma'\nabla'\Phi' = 0
\end{equation*}

\subsubsection*{Analytical solution:}
The analytical solution for the current carrying wire can be derived in terms of variables in potential formulation. A hollow cylinder with inner radius \( r_i \) and outer radius \( r_o \) is considered. The cylinder extends from \( z = -a \) at the bottom surface to \( z = +a \) at the top surface. The top surface is held at a constant potential \( V_o \), while the bottom surface is grounded at zero potential.

The equation for electric potential in a medium with constant electrical conductivity $\sigma$ is given by equation \ref{eqn:13}:

\begin{equation*}
    \nabla \cdot \nabla \Phi = 0
\end{equation*}

Expressing the above Laplace equation in cylindrical coordinates:
\begin{equation*}
	\frac{1}{r} \frac{\partial}{\partial r} \left(r \frac{\partial \Phi}{\partial r}\right) + \frac{1}{r^2} \frac{\partial^2 \Phi}{\partial \theta^2} + \frac{\partial^2 \Phi}{\partial z^2} = 0
\end{equation*}

Since the current flows purely in the z direction, the Laplace equation simplifies to:
\begin{equation*}
\frac{\partial^2 \Phi}{\partial z^2} = 0
\end{equation*}

Integrating twice with respect to $z$ yields $\Phi = C_1 z + C_2$, where $C_1$ and $C_2$ are constants. By applying the boundary conditions, the constants are determined, leading to the final solution:
\begin{equation*}
	\Phi = \frac{V_o}{2a} (z+a)
\end{equation*}

\clearpage
\begin{figure}[H]
	\centering
	\begin{subfigure}[b]{0.4\textwidth}
		\centering
		\includegraphics[width=\textwidth,trim={0.1cm 0.2cm 13cm 2cm},clip]{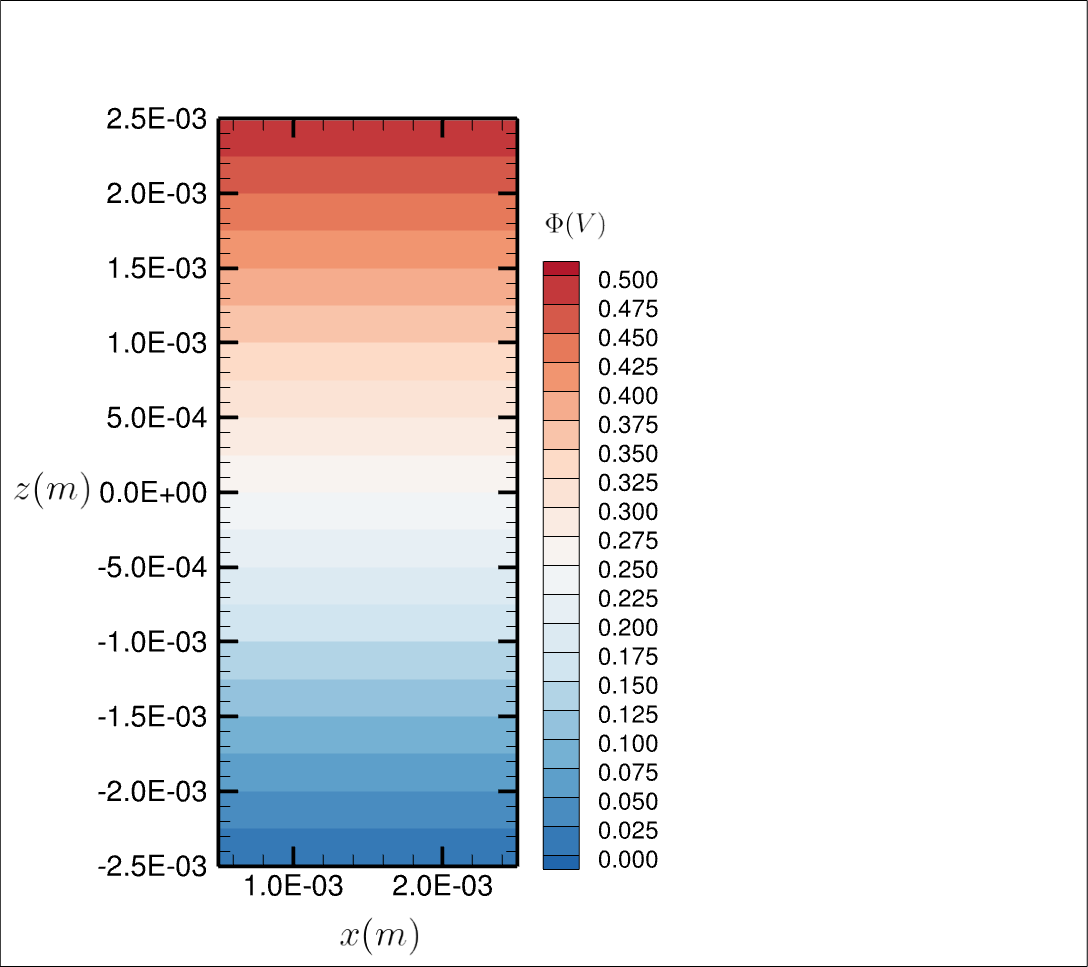}
		\caption{}
	\end{subfigure}
	\hfill
	\begin{subfigure}[b]{0.4\textwidth}
		\centering
		\includegraphics[width=\textwidth,trim={0.2cm 0.2cm 13cm 2cm},clip]{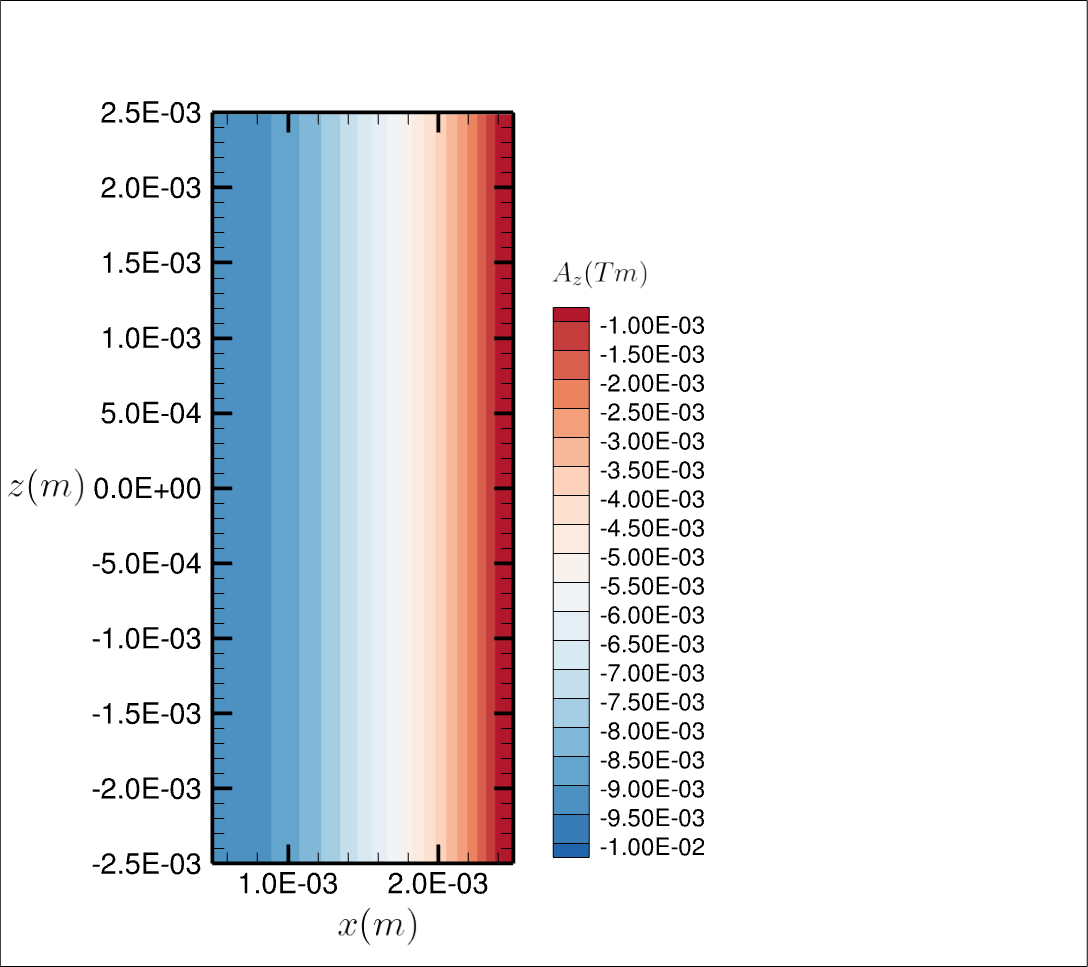}
		\caption{}
	\end{subfigure}
	\hfill
	\begin{subfigure}[b]{0.4\textwidth}
		\centering
		\includegraphics[width=\textwidth,trim={0.1cm 0.1cm 14cm 2cm},clip]{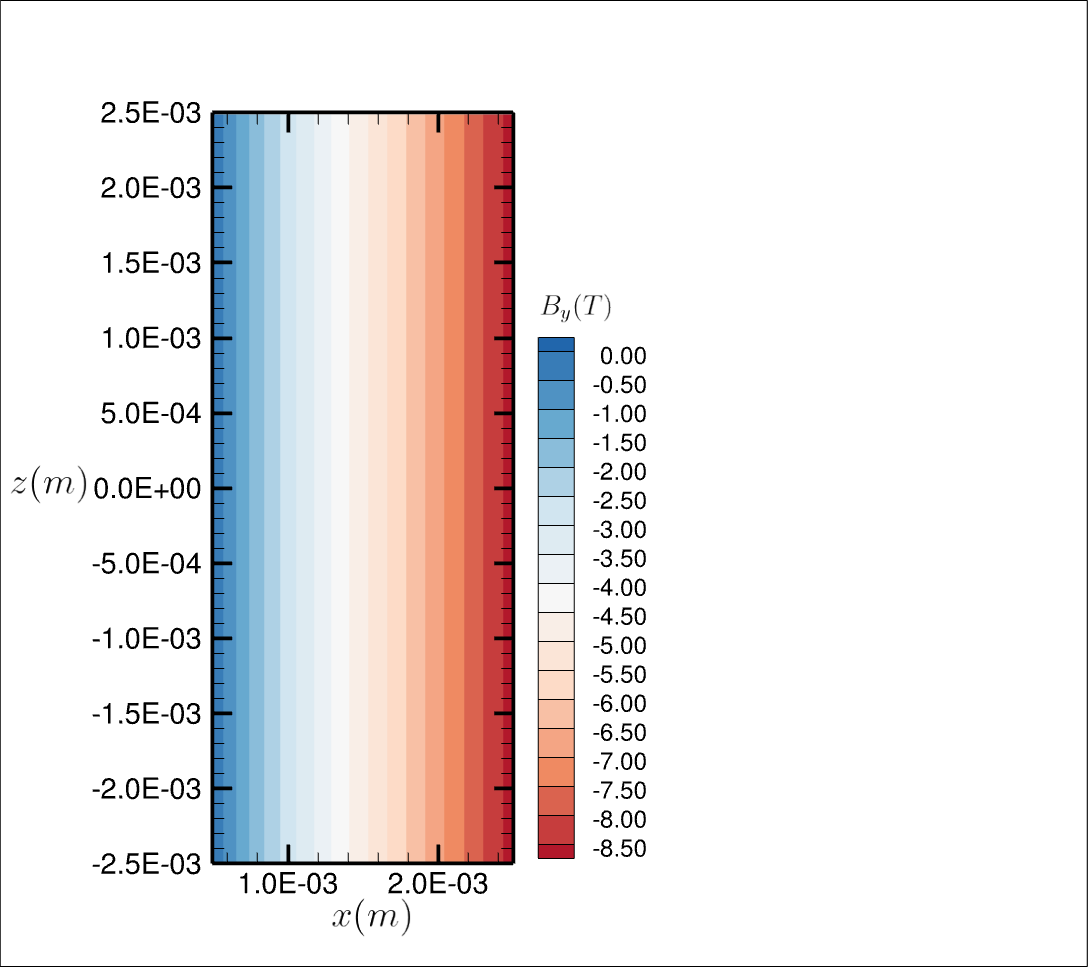}
		\caption{}
	\end{subfigure}
	\hfill
	\begin{subfigure}[b]{0.4\textwidth}
		\centering
		\includegraphics[width=\textwidth,trim={0.1cm 0.1cm 14cm 2cm},clip]{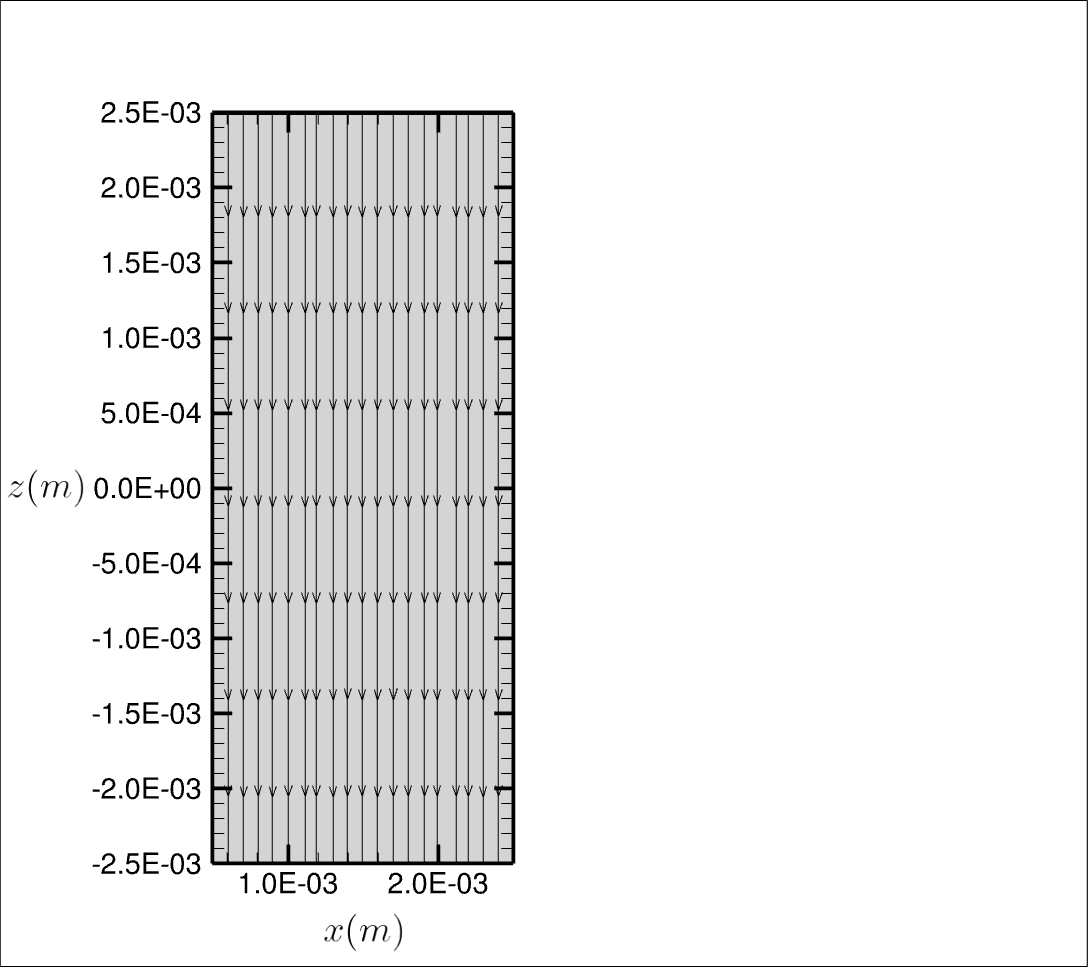}
		\caption{}
	\end{subfigure}
	\hfill
	\caption{ Simulation results of  steady current flow in a wire along the axial direction. The figures display (a) electric potential, (b) magnetic vector potential, (c) magnetic field and (d) streamlines of current density.}
	\label{fig:3.4}
\end{figure}

To find the current density, Ohm's law is applied as $\mathbf{J} = - \sigma \nabla \Phi$. In cylindrical coordinates, the gradient of the electric potential is expressed as:
\begin{equation*}
	\nabla \phi = \hat{r} \frac{\partial \phi}{\partial r} + \hat{\theta} \frac{1}{r} \frac{\partial \phi}{\partial \theta} + \hat{z} \frac{\partial \phi}{\partial z}
\end{equation*}
Given that $\Phi(z)$ is solely a function of $z$, only the axial component is non-vanishing:
\begin{equation*} 
  J_z = -\sigma \frac{\partial \Phi}{\partial z} = -\frac{\sigma V_o}{2a} = -\frac{\sigma V_o}{l}
\end{equation*}

\clearpage

To determine the vector potential, equation \ref{eqn:14} needs to be solved:
\begin{equation*}
	\nabla \cdot \nabla \mathbf{A} = - \mu_0 \mathbf{J}
\end{equation*}

To determine the vector potential, it is necessary to solve equation \ref{eqn:14}, given by $\nabla \cdot \nabla \mathbf{A} = - \mu_0 \mathbf{J}$. Since the current is purely axial, only the axial component of the vector potential, $A_z$, needs to be considered: $\nabla \cdot \nabla A_z = - \mu_0 J_z$. In cylindrical coordinates, the Laplacian of $A_z$ is expressed as:
\begin{equation*}
	\frac{1}{r} \frac{\partial}{\partial r} \left(r \frac{\partial A_z}{\partial r}\right) + \frac{1}{r^2} \frac{\partial^2 A_z}{\partial \theta^2} + \frac{\partial^2 A_z}{\partial z^2} = - \mu_0 J_z
\end{equation*}

Considering the symmetry, the vector potential depends only on the radius, simplifying the equation to:

\begin{equation*}
    \frac{1}{r} \frac{\partial}{\partial r} \left(r \frac{\partial A_z}{\partial r}\right) = - \mu_0 J_z \quad \Longrightarrow \quad \frac{\partial}{\partial r} \left(r \frac{\partial A_z}{\partial r}\right) = - \mu_0 r J_z
\end{equation*}

Integrating the equation twice yields:
\begin{equation*}
    A_z =  - \left(\frac{\mu_0   J_z }{4}\right)  r^2+ C_1 \ln r + C_2
\end{equation*}

\begin{figure}[H]
	\centering
	\begin{subfigure}[b]{0.42\textwidth}
		\centering
		\includegraphics[width=\textwidth]{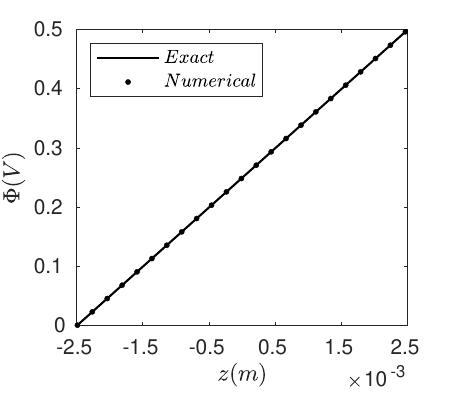}
		\caption{}
	\end{subfigure}
	\hfill
	\begin{subfigure}[b]{0.42\textwidth}
		\centering
		\includegraphics[width=\textwidth]{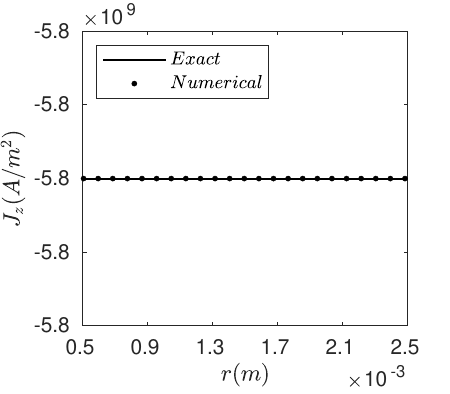}
		\caption{}
	\end{subfigure}
	\hfill
	\begin{subfigure}[b]{0.42\textwidth}
		\centering
		\includegraphics[width=\textwidth]{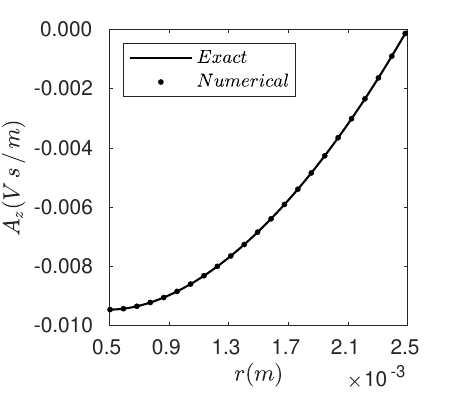}
		\caption{}
	\end{subfigure}
	\hfill
	\begin{subfigure}[b]{0.42\textwidth}
		\centering
		\includegraphics[width=\textwidth]{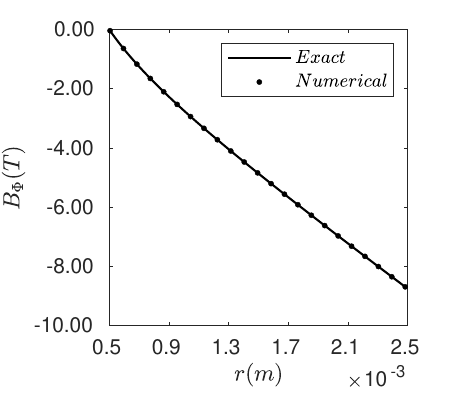}
		\caption{}
	\end{subfigure}
	\hfill
	\caption{ Comparison between the numerical and analytical solutions for a wire carrying steady axial current (a) electric potential, (b) current density, (c) magnetic vector potential  and (d) magnetic field.}
	\label{fig:3.5}
\end{figure}

\clearpage

The constants $C_1$ and $C_2$ must be determined based on boundary conditions. To ensure that the magnetic field is continuous at the inner surface of the wire, the vector potential $\textbf{A}$ must also be continuous across this boundary since $\textbf{B} = \nabla \times \textbf{A}$. This implies that the gradient of the vector potential should be zero at this surface, leading to the boundary condition:
\begin{equation*}
	\frac{\partial \textbf{A}}{\partial r}\bigg|_{r=r_i} = \textbf{0}
\end{equation*}

Furthermore, to confine the magnetic field within the wire and ensure that it decays to zero as we move away from the wire, the vector potential should decay to zero as well. Mathematically, this means that the vector potential should be zero at the outer surface of the wire, leading to the boundary condition:

\begin{equation*}
	\textbf{A}\bigg|_{r=r_o} = \textbf{0}
\end{equation*}

Applying the zero gradient boundary condition at the inner surface and Dirichlet boundary condition at the outer surface yields:

\begin{equation*}	A_z =  \left(\frac{\mu_0  \sigma V_o }{4l}\right) (r^2  - r_o^2) - \left(\frac{\mu_0   \sigma V_o r_i^2 }{2l}\right) ( \ln r - \ln r_o )
\end{equation*}

The magnetic field can be calculated using the curl of the vector potential:
\begin{equation*}
	\mathbf{B} =  \nabla \times \mathbf{A} =\left( \frac{1}{r} \frac{\partial A_z}{\partial \theta} - \frac{\partial A_\theta}{\partial z}\right)\hat{r} + \left(\frac{\partial A_r}{\partial z} - \frac{\partial A_z}{\partial r}\right)\hat{\theta} + \frac{1}{r}\left(\frac{\partial}{\partial r}(rA_\theta) - \frac{\partial A_r}{\partial \theta}\right)\hat{z}
\end{equation*}
The only non-vanishing component is ($A_z$ depends solely on $r$):
\begin{equation*}
B_\theta = - \frac{\partial A_z}{\partial r}  =   - \frac{\mu_0   \sigma V_o }{4a r} (r^2 - r_i^2)
\end{equation*}

\subsection{Steady Radial Current Carrying Wire}

\begin{figure}[H]
	\centering
			\begin{tikzpicture}[scale=1.0]
				
				\tikzset{
					line/.style={draw=blue, thick},
					fillstyle/.style={fill=black, fill opacity=0.2},
					dashedline/.style={draw=blue, thick, dashed}
				}
				
				\draw[line] (0,0) ellipse (2cm and 1cm);
				
				\draw[line] (0,0) ellipse (0.5cm and 0.25cm);
				
				\draw[black, dashed] (0,-0.5) -- (0,-4);
				
				\coordinate (InnerTangentPoint) at (0.5,-0.125);
				\coordinate (OuterTangentPoint) at (1.85,-0.375);
				\coordinate (TopInner) at ($(InnerTangentPoint)+(0,0.25)$);
				\coordinate (TopOuter) at ($(OuterTangentPoint)+(0,0.75)$);
				\coordinate (BottomInner) at (InnerTangentPoint);
				\coordinate (BottomOuter) at (OuterTangentPoint);
				
				\fill[fillstyle] (InnerTangentPoint) -- (OuterTangentPoint) -- (TopOuter) -- (TopInner) -- cycle;
				
				\draw[line] (InnerTangentPoint) -- (TopInner);
				\draw[line] (OuterTangentPoint) -- (TopOuter);
				\draw[line] (TopInner) -- (TopOuter);
				\draw[line] (BottomInner) -- (BottomOuter);
				
				\coordinate (LeftVerticalTangent) at (-2, 0); 
				\draw[line] (LeftVerticalTangent) -- ++(0, -4); 
				\coordinate (RightVerticalTangent) at (2, 0); 
				\draw[line] (RightVerticalTangent) -- ++(0, -4); 
				
				\draw[line] (-2,-4) arc (180:360:2cm and 1cm); 
				\draw[dashedline] (-2,-4) arc (180:0:2cm and 1cm); 
				
				\tikzset{currentarrow/.style={->, >=latex, line width=0.4mm, red}} 
				
				\newcommand{\currentarrowone}{
					\draw[currentarrow] (0.75,0) -- (1.75,0);
					\node at (0.0,0.0) {$J$};
				}
				
				\newcommand{\currentarrowtwo}{
					\draw[currentarrow] (0,0.3) -- (0,0.9);
				}
				
				\newcommand{\currentarrowthree}{
					\draw[currentarrow] (-0.75,0) -- (-1.75,0);
				}
			
				\newcommand{\currentarrowfour}{
				\draw[currentarrow] (0,-0.3) -- (0,-0.9);
				}
				
				\currentarrowone
				\currentarrowtwo
				\currentarrowthree
				\currentarrowfour
		
			\end{tikzpicture}

	\caption{Hollow cylindrical geometry with a steady radial current. Boundary labels and the representative computational mesh is presented in  Figure \ref{fig:2}.}
 \label{fig:3.6}
\end{figure}
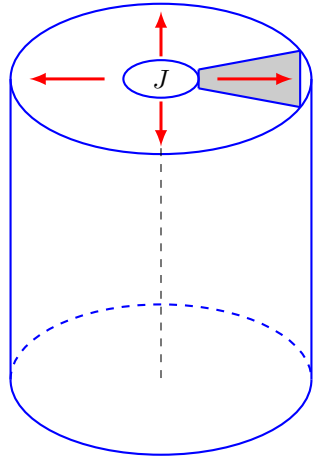

A hollow cylinder is subjected to a constant potential difference between its inner and outer cylindrical surfaces, as illustrated in Figure \ref{fig:3.6}. The cylinder has a length of $l$ and inner and outer radii of $r_i$ and $r_o$, respectively. Various parameters of the cylinder are computed using numerical simulations, including the potential distribution along its radius ($\Phi$), the surface current density ($\mathbf{J}$) inside the cylinder, the magnetic vector potential ($\mathbf{A}$) and the magnetic field inside the cylinder ($\mathbf{B}$). 

\clearpage

\begin{table}[H]
    \centering
	\caption{Geometry, physical property and boundary condition parameters used for the numerical simulation of  hollow cylinder carrying a steady current in the radial direction.}
    \begin{tabular}{llll}
        \toprule
        Parameter & Description & Value & Units \\
        \midrule
		$r_i$ & Inner radius of cylinder & $0.1$ & m \\
		$r_o$ & Outer radius of cylinder & $0.15$ & m \\
		$H$ & Height of cylinder & $0.01$ & m \\
		$\mu_0$ & Magnetic permeability & $4 \pi \times 10^{-7}$ & H/m \\
		$\sigma$ & Electrical conductivity & $5.8 \times 10^7$ & S/m \\
		$\Phi_b$ & Potential at back boundary & $0.01$ & V \\
		$\Phi_f$ & Potential at front boundary & $0.0$ & V \\
		$\mathbf{A}_t, \, \mathbf{A}_b$ & Vector potential at top and bottom boundary & $0.0$ & T $\cdot$ m \\
        \bottomrule
    \end{tabular}
	\label{tab:3.5}
\end{table}

\begin{table}[H]
    \centering
	\caption{Boundary conditions used for the numerical simulation of  hollow cylinder carrying a steady current in the radial direction.}
    \begin{tabular}{lll}
        \toprule
        & \multicolumn{2}{c}{Boundary Condition}\\
        \cmidrule(r){2-3}
        Boundary & Potential ($\Phi$) & Vector potential ($\mathbf{A}$) \\
        \midrule
		Top boundary & Zero normal gradient & Fixed value \\
		Bottom boundary & Zero normal gradient & Fixed value \\
        Left boundary & Zero normal gradient & Zero normal gradient \\
		Right boundary & Zero normal gradient & Zero normal gradient \\
		Front boundary & Fixed value & Zero normal gradient \\
		Back boundary & Fixed value & Zero normal gradient \\
        \bottomrule
    \end{tabular}
	\label{tab:3.6}
\end{table}

\subsubsection*{Analytical solution:}

For the case of a hollow cylinder with inner radius $r_i$ and outer radius $r_o$, and with the outer surface at zero potential and the inner surface at  potential $V_o$, the analytical solution is derived as follows:

Following similar approach to the axial current case, the electric potential can be found using the Laplace equation in cylindrical coordinates.

\begin{equation*}
	\frac{1}{r} \frac{\partial}{\partial r} \left(r \frac{\partial \Phi}{\partial r}\right) + \frac{1}{r^2} \frac{\partial^2 \Phi}{\partial \theta^2} + \frac{\partial^2 \Phi}{\partial z^2} = 0
\end{equation*}

Since the current is purely along the  $r$ direction, the Laplace equation simplifies to:
\begin{equation*}
	 \frac{\partial}{\partial r} \left(r \frac{\partial \Phi}{\partial r}\right)  = 0
\end{equation*}

Integrating twice with respect to $r$ yields $\Phi = C_1 \ln r + C_2$, where $C_1$ and $C_2$ are constants. Applying the boundary conditions, the electric potential is obtained as follows:
\begin{equation*}
	\Phi = \frac{V_o}{\ln (r_i/r_o)} \ln (r/r_o)
\end{equation*}

To calculate the current density, Ohm's law can be used as before $\mathbf{J} = - \sigma \nabla \Phi$. 
In cylindrical coordinates, the gradient of the potential is given by:
\begin{equation*}
	\nabla \phi = \hat{r} \frac{\partial \phi}{\partial r} + \hat{\theta} \frac{1}{r} \frac{\partial \phi}{\partial \theta} + \hat{z} \frac{\partial \phi}{\partial z}
\end{equation*}

Since $\Phi$ is only a function of $r$: $ \displaystyle J(r) = - \sigma \frac{\partial \Phi}{\partial r} \hat{r}$. Substituting the previously derived expression for $\phi$, the result is:
\begin{equation*}
    J(r) = - \frac{\sigma V_o}{r \ln (r_i/r_o)}
\end{equation*}

\clearpage

\begin{figure}[H]
	\centering
	\begin{subfigure}[b]{0.7\textwidth}
		\centering
		\includegraphics[width=1.0\textwidth,trim={0.2cm 1cm 2cm 20cm},clip]{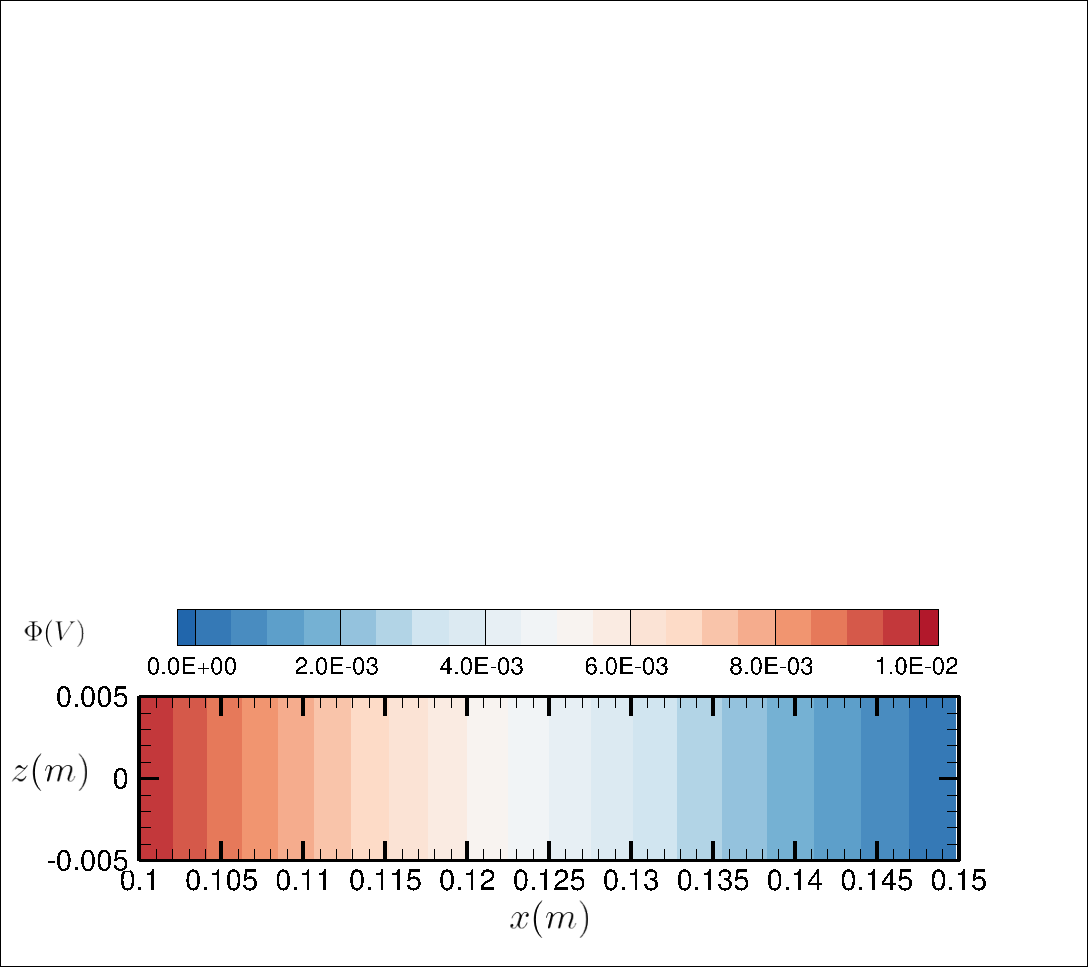}
		\caption{}
	\end{subfigure}
	\hfill
	\begin{subfigure}[b]{0.7\textwidth}
		\centering
		\includegraphics[width=1.0\textwidth,trim={0.2cm 1cm 3cm 20cm},clip]{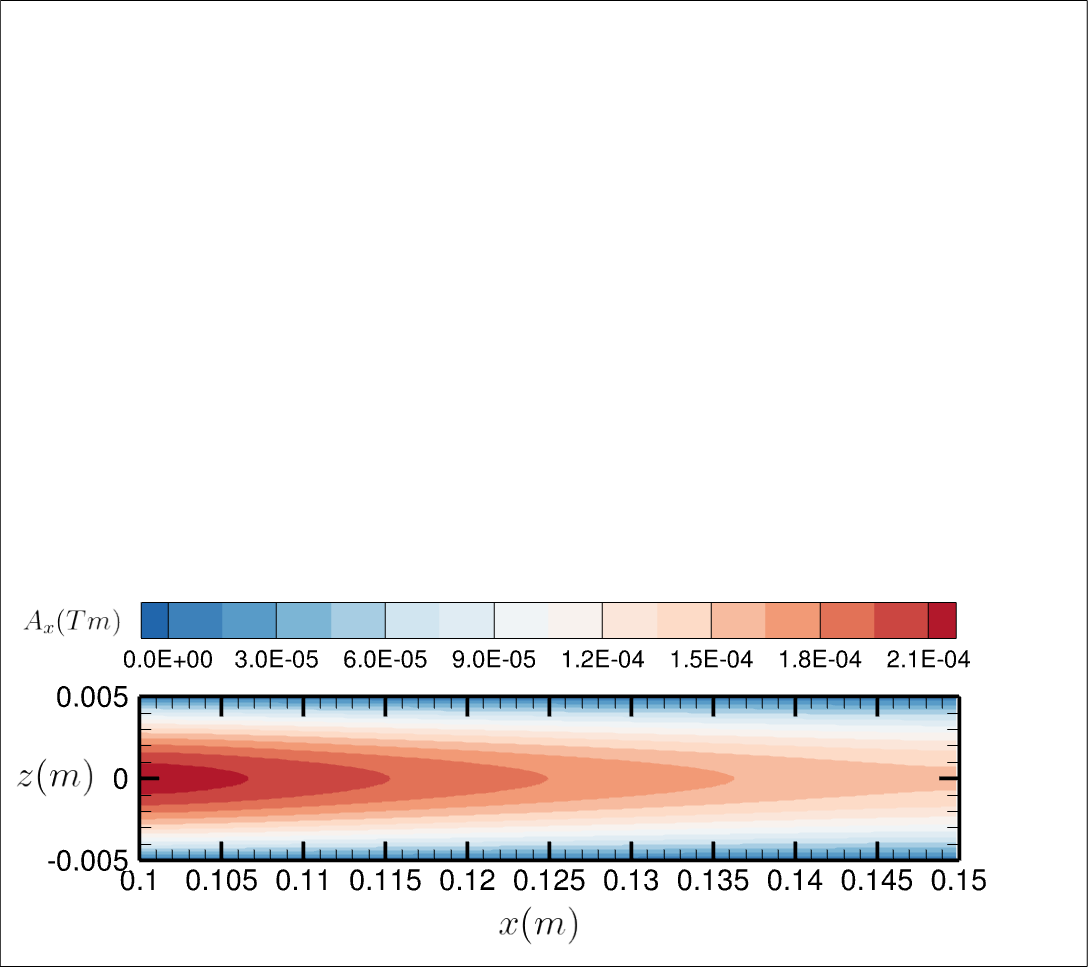}
		\caption{}
	\end{subfigure}
	\hfill
	\begin{subfigure}[b]{0.7\textwidth}
		\centering
		\includegraphics[width=1.0\textwidth,trim={0.2cm 1cm 3cm 20cm},clip]{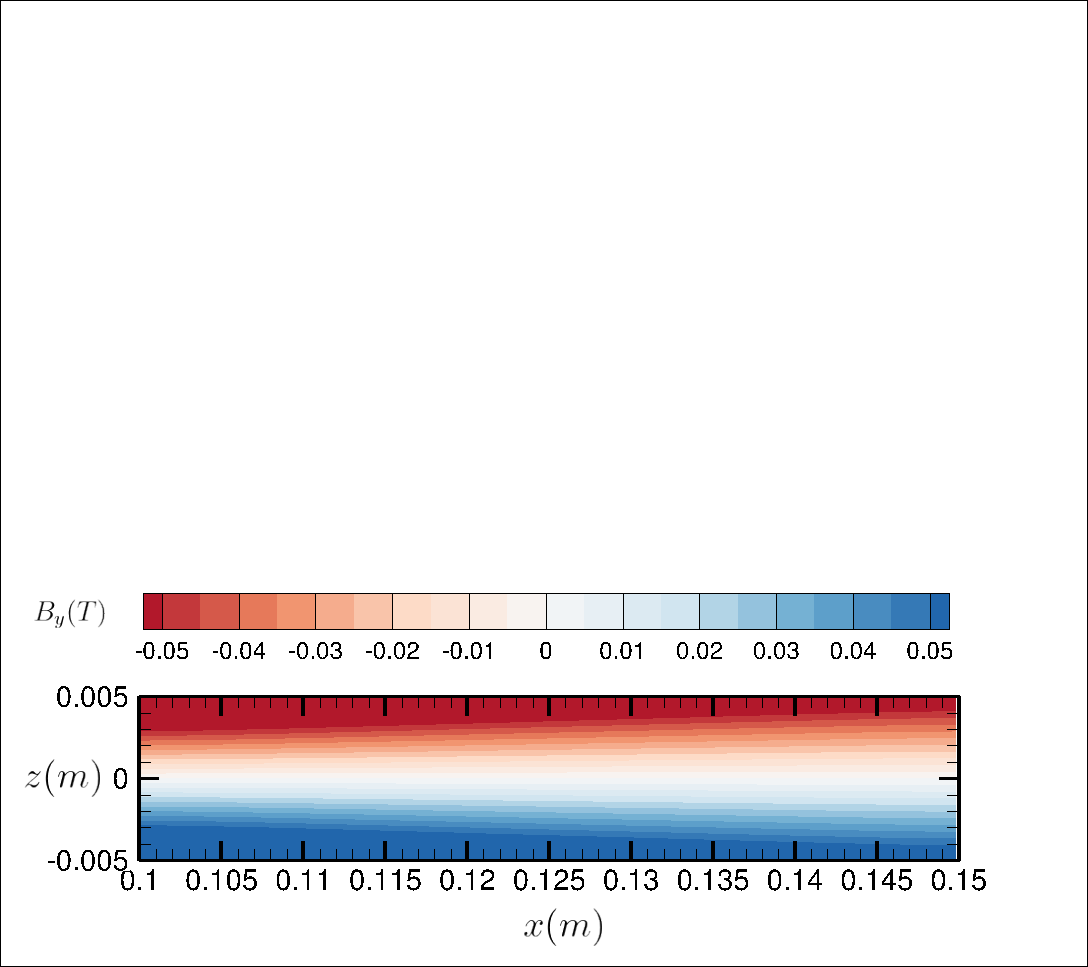}
		\caption{}
	\end{subfigure}
	\hfill
	\begin{subfigure}[b]{0.7\textwidth}
		\centering
		\includegraphics[width=1.0\textwidth,trim={0.2cm 1cm 3cm 20cm},clip]{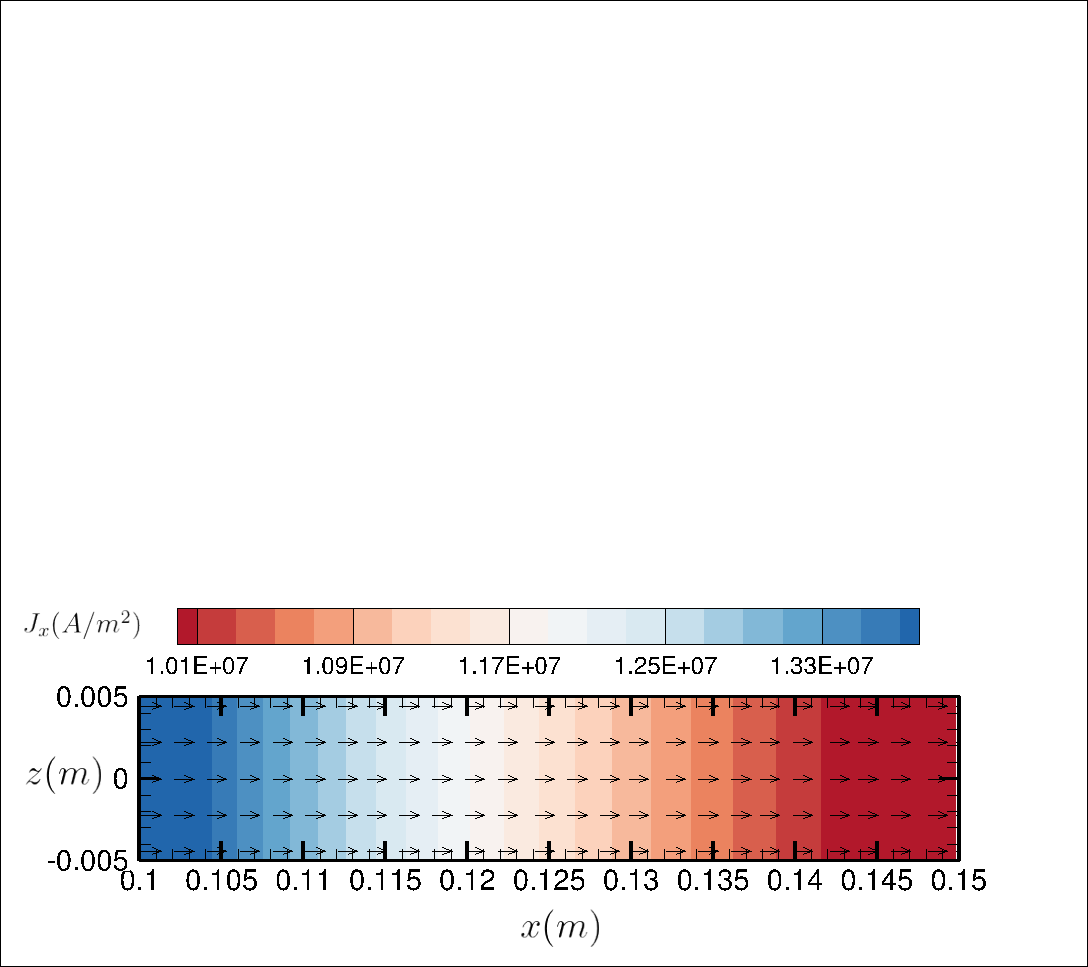}
		\caption{}
	\end{subfigure}
	\hfill
	\caption{ Simulation results of  steady current flow in a hollow cylinder along the radial direction. The figures display (a) the electric potential, (b) the magnetic vector potential, (c) the magnetic field, and (d) the current density.}
	\label{fig:3.12}
\end{figure}

Vector potential can be determined by solving $\nabla \cdot \nabla \textbf{A} = - \mu_0  \textbf{J}_{\Phi}$, where $\textbf{J}_{\Phi}  = - \sigma \nabla \Phi$.

Since the current is purely radial, only the radial component of the vector potential is considered : $\nabla \cdot \nabla A_r = - \mu_0  J(r)$. 

\clearpage

\begin{figure}[H]
	\centering
	\begin{subfigure}[b]{0.45\textwidth}
		\centering
		\includegraphics[width=1.2\textwidth]{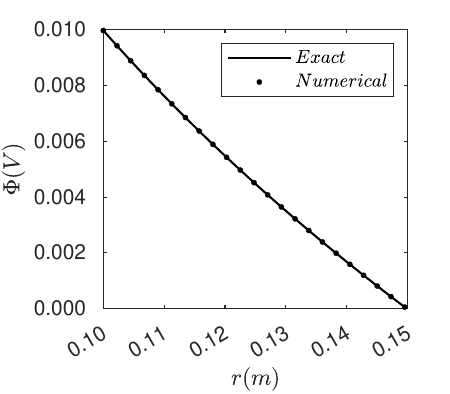}
		\caption{}
	\end{subfigure}
	\hfill
	\begin{subfigure}[b]{0.45\textwidth}
		\centering
		\includegraphics[width=1.2\textwidth]{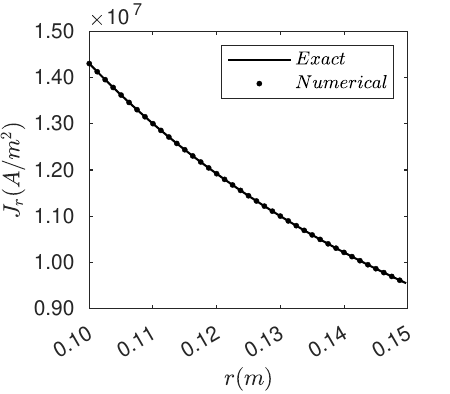}
		\caption{}
	\end{subfigure}
	\hfill
	\begin{subfigure}[b]{0.45\textwidth}
		\centering
		\includegraphics[width=1.2\textwidth]{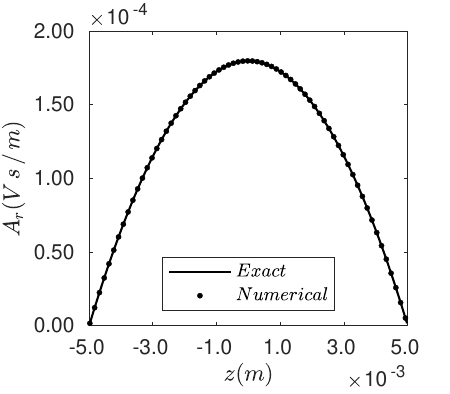}
		\caption{}
	\end{subfigure}
	\hfill
	\begin{subfigure}[b]{0.45\textwidth}
		\centering
		\includegraphics[width=1.2\textwidth]{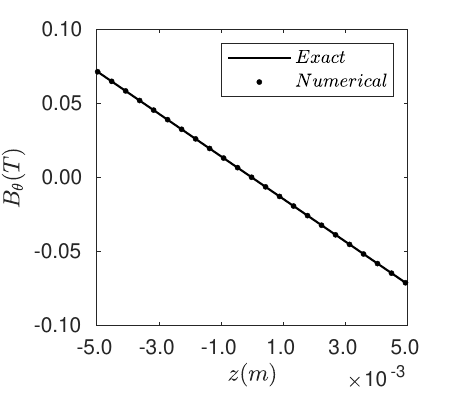}
		\caption{}
	\end{subfigure}
	\hfill
	\caption{ Comparison between the numerical and exact solutions for a hollow cylinder carrying steady radial current (a) the electric potential, (b) the current density, (c) the magnetic vector potential and (d) the magnetic field}
	\label{fig:3.13}
\end{figure}

In cylindrical coordinates, the Laplacian can be expressed as:
\begin{equation*}
	\frac{1}{r} \frac{\partial}{\partial r} \left(r \frac{\partial A_r}{\partial r}\right) + \frac{1}{r^2} \frac{\partial^2 A_r}{\partial \theta^2} + \frac{\partial^2 A_r}{\partial z^2} = - \mu_0  J(r)
\end{equation*}

The central part of the cylinder far from inner and outer radius is considered. In this region, radial dependence can be neglected. Therefore, the equation simplifies to:
\begin{equation*}
	 \frac{\partial^2 A_r}{\partial z^2} = - \mu_0  J(r)
\end{equation*}

Integrating twice with respect to z, $ \displaystyle A_r = -\frac{1}{2} \mu_0 J(r) z^2 + C_1z + C_2$. Here, $C_1$ and $C_2$ are integration constants, which can be determined using the boundary conditions $A_r(r,\theta,a)=A_r(r,\theta,-a)=0$. 

Consequently, the resulting expression is:
\begin{equation*}
	  A_r(r, \theta, z)  = \frac{1}{2} \mu_0 J(r)  (a^2 - z^2)
\end{equation*}
The magnetic field can be calculated using the curl of the vector potential, which is given by:
\begin{equation*}
	\mathbf{B} =  \nabla \times \mathbf{A} =\left( \frac{1}{r} \frac{\partial A_z}{\partial \theta} - \frac{\partial A_\theta}{\partial z}\right)\hat{r} + \left(\frac{\partial A_r}{\partial z} - \frac{\partial A_z}{\partial r}\right)\hat{\theta} + \frac{1}{r}\left(\frac{\partial}{\partial r}(rA_\theta) - \frac{\partial A_r}{\partial \theta}\right)\hat{z}
\end{equation*}
Only non-vanishing component is 
\begin{equation*}
	\mathbf{B} = \nabla \times \mathbf{A} =  \frac{\partial A_r}{\partial z} \hat{\theta} = -\mu_0 J(r) z \hat{\theta} = \frac{\mu_0  \sigma V_o z}{r \ln (r_i/r_o)}  \hat{\theta}
\end{equation*}

\subsection{Axisymmetric Electrically driven Flow in Annular Channel}

The simulation setup consists of two coaxial cylinders filled with liquid metal \cite{suponitsky2020axisymmetric,suponitsky2020test}. A uniform axial magnetic field, referred to as $B_z$=$B_o$, permeates the liquid metal. When an electric current $I_o$ is applied from the inner to the outer cylinder through the liquid metal medium, it triggers the metal to spin in azimuthal direction. The cylinder has a length of $l$ and inner and outer radii of $r_i$ and $r_o$ respectively. Angular momentum along the $z$ axis, away from the side walls, is compared with the exact solutions in literature \cite{khalzov2008equilibrium,suponitsky2022magnetohydrodynamics}.

\begin{figure}[htbp]
	\centering
	\begin{tikzpicture}[scale=1.2, >=Stealth] 
		\usetikzlibrary{patterns} 
		
		\coordinate (A) at (0,0);
		\coordinate (B) at (5,0);
		\coordinate (C) at (5,3);
		\coordinate (D) at (0,3);
		\def\wallthickness{0.2} 
		
		\fill[pattern=north east lines, pattern color=black] (A) rectangle ($(A)+(-\wallthickness,3)$); 
		\fill[pattern=north east lines, pattern color=black] (A) rectangle ($(B)+(0,-\wallthickness)$); 
		\fill[pattern=north east lines, pattern color=black] (B) rectangle ($(C)+(\wallthickness,0)$); 
		\fill[pattern=north east lines, pattern color=black] (D) rectangle ($(C)+(0,\wallthickness)$); 
		
		\draw[line width=2pt] (A) -- (B) -- (C) -- (D) -- cycle;
		
		\draw[->, line width=2pt] (2,1.5) -- ++(1,0) node[midway,below] {\(I_x\)}; 
		
		\draw[->, line width=2pt] (5.5,1) -- ++(0,1) node[midway,right] {\(B_z\)}; 
		
		\draw[->, line width=1pt] (0.5,0.5) -- ++(0.5,0) node[right] {\(x\)};
		\draw[->, line width=1pt] (0.5,0.5) -- ++(0,0.5) node[above] {\(z\)};
		
	\end{tikzpicture}
	\caption{Sketch of computational domain used for numerical simulation of axisymmetric electrically driven flow in an annular channel. The picture is a cross section of an annular cylinder. The left and right surfaces correspond to the inner and outer radii of the cylinder. Boundary labels and the representative computational mesh is presented in  Figure \ref{fig:2}.}
	\label{fig:4.6}
\end{figure}
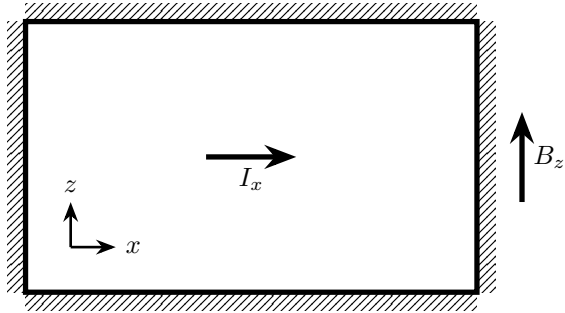

\begin{table}[H]
    \centering
	\caption{Geometry, physical property and boundary condition parameters used for numerical simulation of axisymmetric electrically driven flow in annular channel. All the parameters used are non-dimensional.}
    \begin{tabular}{lll}
        \toprule
        Parameter & Description & Value  \\
        \midrule
    	$r_i$ & Inner radius of cylinder & $1$  \\
    	$r_o$ & Outer radius of cylinder & $5$  \\
    	$H$ & Height of cylinder & $2.0$  \\
    	$\sigma$ & Electrical conductivity &  $1.0$  \\
    	$\Phi_b$ & Potential at back boundary & $ \displaystyle \frac{\ln (r_o/r_i)}{ 2 \pi H \sigma }$ \\
    	$\Phi_f$ & Potential at front boundary & $0.0$  \\
    	$\mathbf{A}_{t/b}$ & Vector potential at top and bottom boundary & $0.0$  \\
    	$B_z$ & Imposed External Magnetic Field & $1.0$ \\
    	$\rho$ & Density of the fluid & $1.0$  \\
    	$p$ & Pressure & $1.0$\\
    	$ts$ & Total simulation time & $1.0$  \\
    	$Re$ & Reynolds number & $1.0$  \\
    	$Ha$ & Hartmann number & $0.1$  \\
        \bottomrule
    \end{tabular}
	\label{tab:4.3}
\end{table}

\clearpage

\begin{table}[H]
    \centering
	\caption{Boundary conditions used for numerical simulation of axisymmetric electrically driven flow in annular channel.}
    \begin{tabular}{llll}
        \toprule
        & \multicolumn{2}{c}{Boundary Condition}\\
        \cmidrule(r){2-4}
        Boundary & Velocity ($ \mathbf{U}$) & Pressure  ($p$)  & Potential  ($\Phi$) \\
        \midrule
    	Top boundary & Fixed value & Zero normal gradient & Zero normal gradient\\
    	Bottom boundary & Fixed value & Zero normal gradient & Zero normal gradient\\
    	Left boundary  & Zero normal gradient & Zero normal gradient & Zero normal gradient\\
    	Right boundary & Zero normal gradient & Zero normal gradient & Zero normal gradient\\
    	Front boundary & Fixed value & Zero normal gradient & Fixed value\\
    	Back boundary & Fixed value & Zero normal gradient & Fixed value\\
        \bottomrule
    \end{tabular}
	\label{tab:4.4}
\end{table}

\begin{figure}[H]
	\centering
	\begin{subfigure}[b]{0.9\textwidth}
		\centering
		\includegraphics[width=0.8\textwidth,trim={0.5cm 1cm 3cm 12cm},clip]{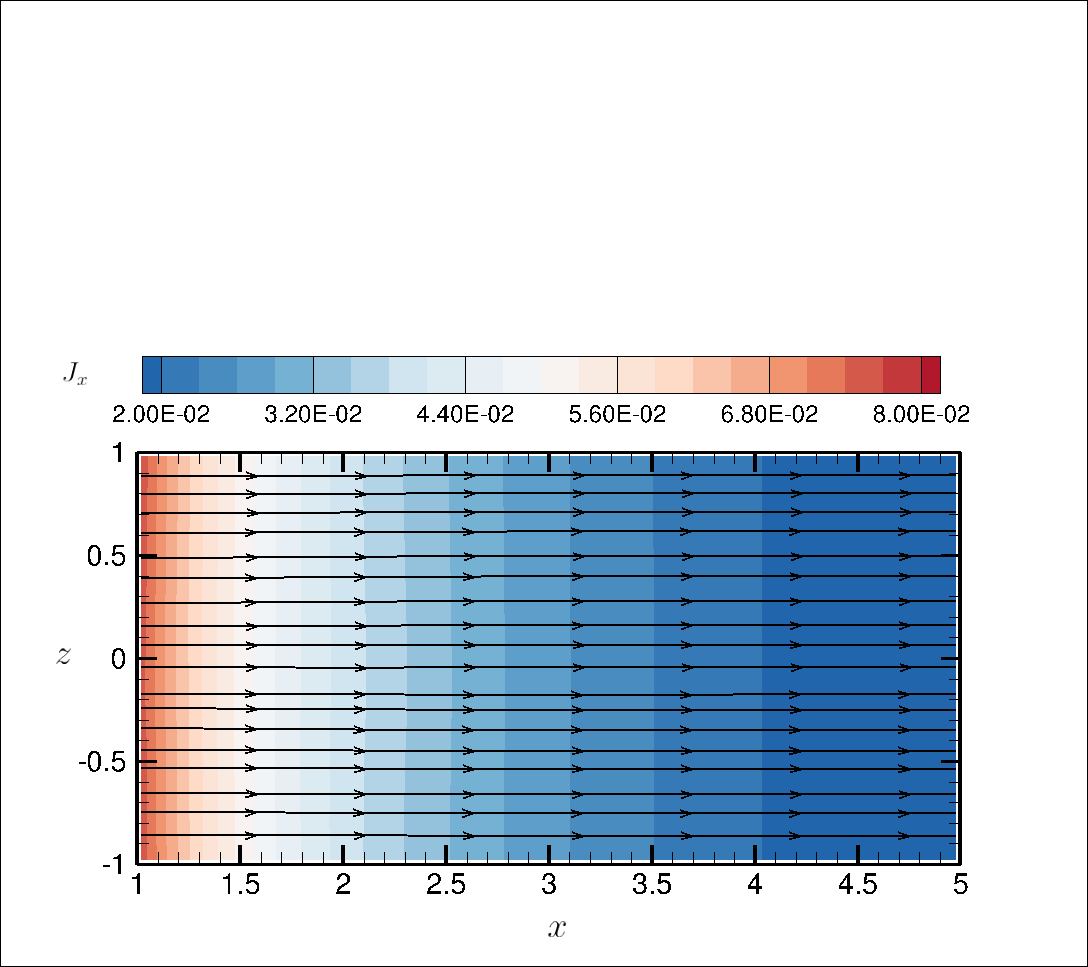}
		\caption{}
    \end{subfigure}
    \hfill
  	\begin{subfigure}[b]{0.9\textwidth}
		\centering
		\includegraphics[width=0.8\textwidth,trim={0.1cm 0.2cm 3cm 12cm},clip]{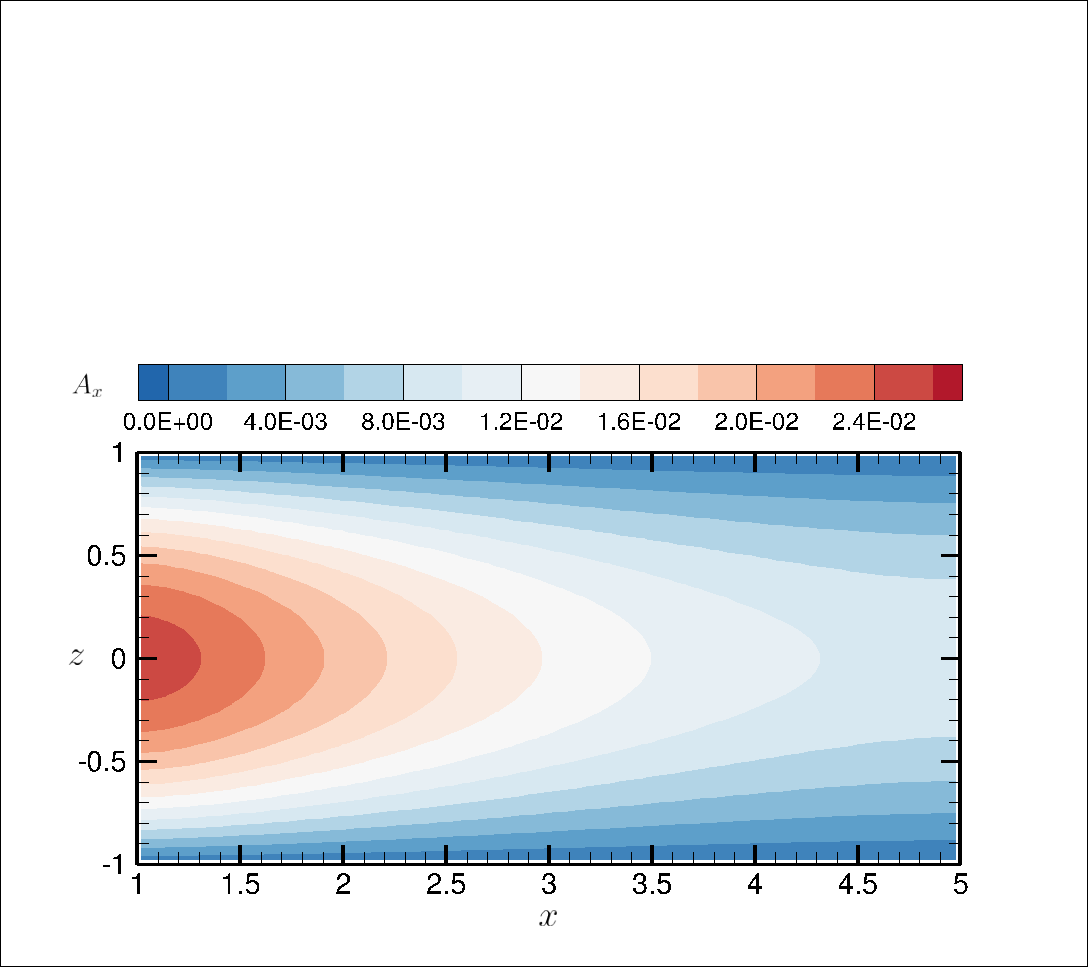}
		\caption{}
	\end{subfigure}
	\caption{ Simulation results of  steady electrically driven flow in annular channel. The figures display (a) stream lines of current density coloured with $x$ component of current density, (b) magnetic vector potential. The variables presented in the plot are non-dimensional.}
	\label{fig:4.9}
\end{figure}

\clearpage

\begin{figure}[H]
	\centering
	\begin{subfigure}[b]{0.9\textwidth}
		\centering
		\includegraphics[width=0.9\textwidth,trim={0.1cm 0.2cm 3cm 12cm},clip]{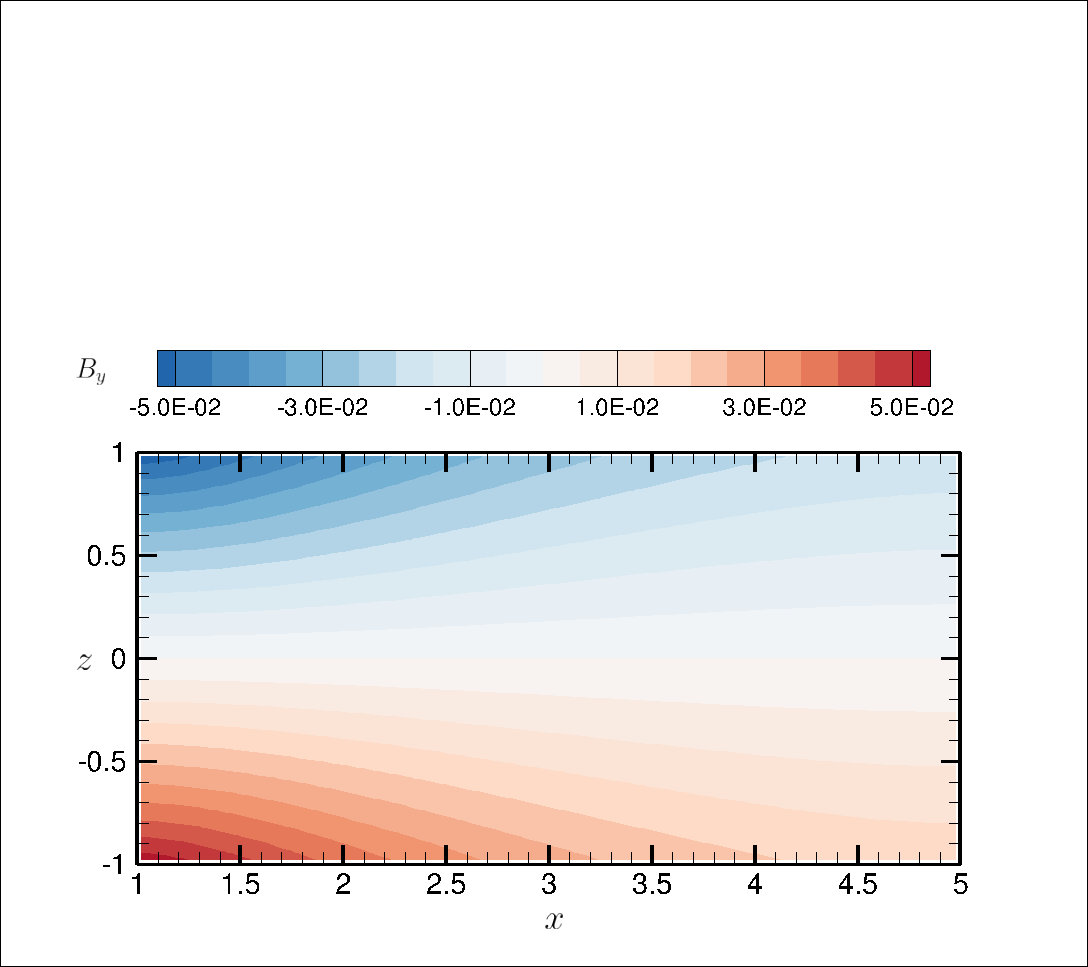}
		\caption{}
	\end{subfigure}
	\hfill
 	\begin{subfigure}[b]{0.9\textwidth}
		\centering
		\includegraphics[width=0.9\textwidth,trim={0.1cm 0.5cm 3cm 12cm},clip]{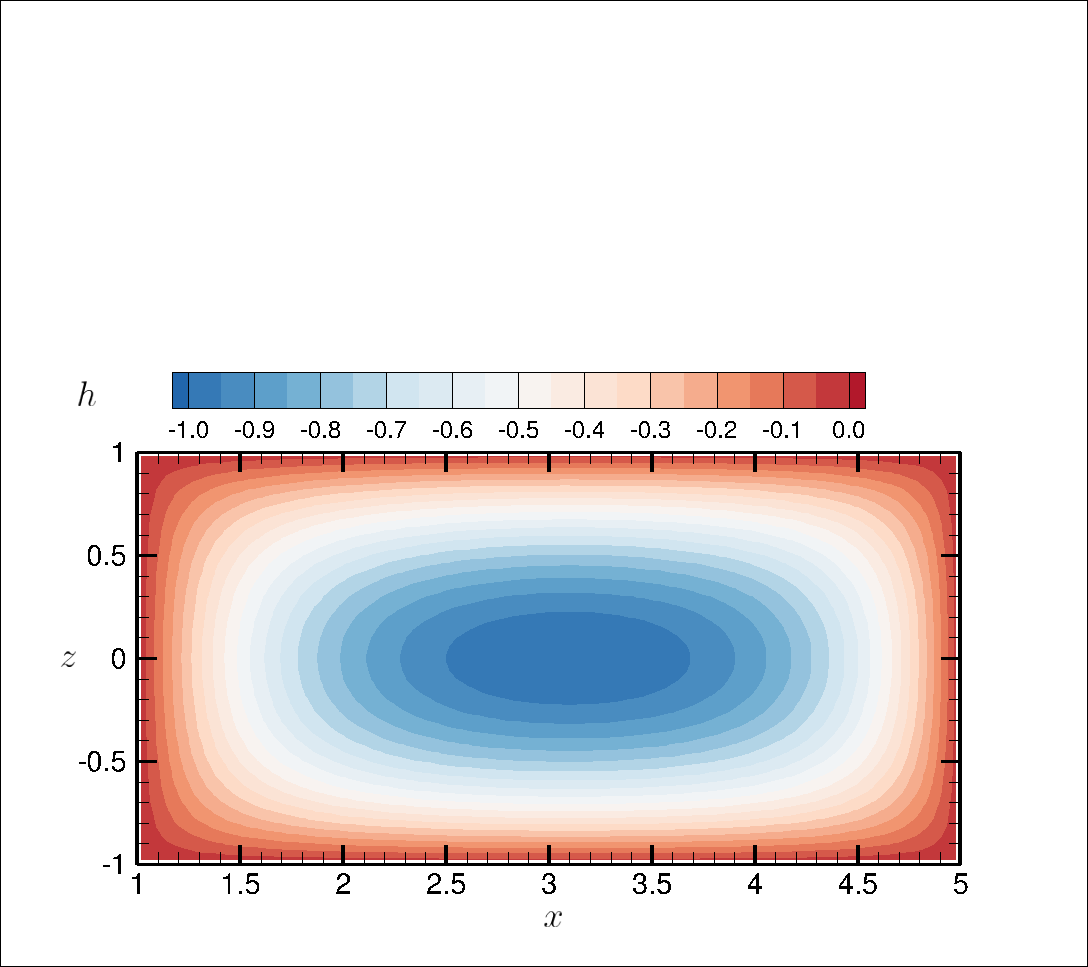}
		\caption{}
	\end{subfigure}
	\caption{ Simulation results of  steady electrically driven flow in annular channel. The figures display (a) induced magnetic field due to applied current, (b) the angular momentum of the rotating fluid ($h = x \, U_y = r \, U_\theta$). The variables presented in the plot are non-dimensional.}
	\label{fig:4.10}
\end{figure}

The governing Maxwell-Navier-Stokes equations are \cite{murugaiyan2024modeling}

	\begin{equation*}
		\nabla \cdot \sigma \nabla \Phi  -
		\nabla \cdot \left( \sigma \textbf{U} \times \textbf{B}\right)
		= 0
	\end{equation*}
	\begin{equation*}
		\nabla \cdot \textbf{U}   = 0
	\end{equation*}
	\begin{equation*}
		\frac{\partial \rho \textbf{U}}{\partial t} +
		\nabla \cdot \left(\rho \textbf{U} \textbf{U} \right) - 
		\frac{1}{Re} \nabla \cdot \left(  \nabla \textbf{U} \right)+ 
		\nabla p -
		\frac{Ha^2  }{ Re} \, \mathbf{J} \times \mathbf{B} = 0
	\end{equation*}

 \clearpage
 
The non-dimensional form of the Maxwell equations is derived by introducing 
length scale $L_o$,
electric conductivity scale $\sigma_o$,
electric potential scale $\Phi_o$,
velocity scale $U_o$,
magnetic field scale $B_o$.
Note that the scale factor for magnetic field is based on externally applied magnetic field strength and by
using the definition $\mathbf{B} = \nabla \times \mathbf{A}$, vector potential scale can be written as $\displaystyle A_o  = B_o \, L_o$. The non-dimensional form of the Navier stokes equations \cite{batchelor1967introduction,white2008fluid}  is derived by introducing 
density scale $\rho_o$,
time scale $t_o$,
length scale $L_o$,
velocity scale $U_o$ and
pressure scale $p_o$.  The external magnetic field $\mathbf{B}$ is constant.  The Hartmann number ($Ha$) and Reynolds number ($Re$) are defined as $ \displaystyle Ha = B_o L_o \sqrt{\frac{\sigma_o}{\mu}}$ and $ \displaystyle Re = \frac{U_o \rho_o L_o }{ \mu }$, respectively. Here, $\mu$ is the fluid viscosity.

 \begin{figure}[H]
	\centering
	\begin{subfigure}[b]{0.46\textwidth}
		\centering
		\includegraphics[width=\textwidth]{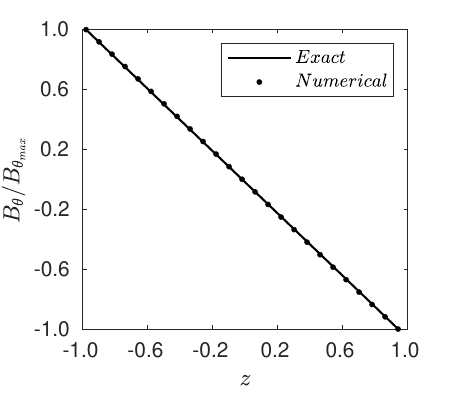}
		\caption{}
	\end{subfigure}
	\hfill
	\begin{subfigure}[b]{0.46\textwidth}
		\centering
		\includegraphics[width=\textwidth]{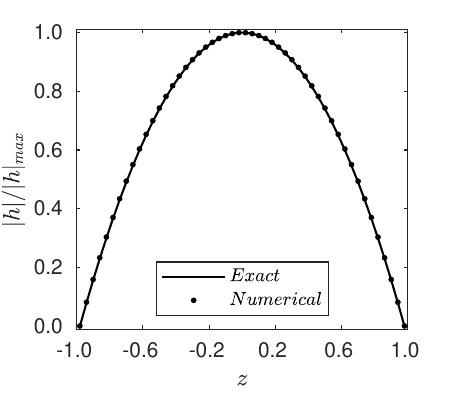}
		\caption{}
	\end{subfigure}
	\hfill
	\caption{ Comparison between the numerical and exact solutions \cite{suponitsky2020axisymmetric,suponitsky2020test,khalzov2008equilibrium} for  steady electrically driven flow in annular channel . The figures display (a) normalized magnetic field due to current, (b) normalized absolute values of angular momentum. The variables in the plot are presented in non-dimensional terms.}
	\label{fig:4.16}
\end{figure}

\paragraph{Algorithm for Solving the Maxwell-Navier-Stokes Equation for Electrically Driven Flows}

\begin{enumerate}

    \item \textbf{Solve the Maxwell Equation}: Calculate the electric potential by solving the Poisson equation as follows:
		\begin{equation*}
			\nabla \cdot \sigma \nabla \Phi  - 
			\nabla \cdot \left( \sigma \textbf{U} \times \textbf{B}\right) = 0
		\end{equation*}
  In this equation, the term $\nabla \cdot \left( \sigma \textbf{U} \times \textbf{B}\right) $ is an explicit source term.

  		\item \textbf{Momentum Predictor}: Estimate the velocity field $\mathbf{U}$ by solving the momentum equation. The pressure gradient term $\nabla p$ and the Lorentz force term $\mathbf{J} \times \mathbf{B}$ are incorporated as explicit source terms based on the initial conditions or values from the previous time step.
		\item \textbf{Construct the H Operator}: The discretized momentum equation can be expressed as follows:
		\begin{equation*}
			a_P \textbf{U}_P + \sum_{N=1}^{nfaces} a_N \textbf{U}_N = b_P
		\end{equation*}
		Considering Rhie-Chow interpolation \cite{jasak1996error,rhie1983numerical,yi2016improved} , the momentum equation is discretized excluding the pressure gradient.
		\begin{equation*}
			a_P \textbf{U}_P + \sum_{N=1}^{nfaces} a_N \textbf{U}_N = b_P - \nabla p 
		\end{equation*}
          
		rewriting as :  \qquad \qquad \qquad \qquad \qquad $a_P \textbf{U}_P  =  \textbf{H}[\textbf{U}] - \nabla p $

		where:
		\begin{equation*}
			\textbf{H}[\textbf{U}] =  b_P - \sum_{N=1}^{nfaces} a_N \textbf{U}_N 
		\end{equation*}

    \clearpage
    
    \item \textbf{Construct and Solve the Pressure Equation}: The velocity obtained from the predictor step is given by the following equation:
    \begin{equation*}
    \textbf{U} =   \frac{1}{a_P} \textbf{H}[\textbf{U}] - \frac{1}{a_P} \nabla p 
    \end{equation*}
    
    Taking divergence of the equation  and then setting the resultant expression to zero
    
    \begin{equation*}  
     \nabla \cdot \left(\frac{1}{a_P} \nabla p  \right) = \nabla \cdot \left( \frac{1}{a_P} \textbf{H}[\textbf{U}]  \right) 
    \end{equation*}

    This leads to a Poisson equation for the pressure. Solving this equation provides the pressure field necessary to correct the velocity field.
    
    \item \textbf{Correct the Velocity Field}: Update the velocity field using the pressure field from the last step:
    \begin{equation*}
    \textbf{U} =   \frac{1}{a_{P}} \textbf{H}[\textbf{U}] - \frac{1}{a_{P}} \nabla p
    \end{equation*}
    
    \item \textbf{Correct the Velocity Flux}: Compute the corrected velocity flux at the cell faces using Rhie-Chow interpolation:
    
    \begin{equation*}
    \mathbf{U}\cdot \mathbf{S}_f =  
    \left(\frac{1}{a_{P}} \mathbf{H}[\mathbf{U}]\right)\cdot \mathbf{S}_f- 
    \left(\frac{1}{a_{P}} \nabla p\right)\cdot \mathbf{S}_f
    \end{equation*}
    Note that this is the velocity flux used in the discretization of divergence term discussed in section \ref{subsec:4.4}. Iterate steps 3 to 6 until convergence is achieved. Once convergence is achieved, move on to the next time step and repeat the process starting from step 1.
\end{enumerate}

\section{Free Surface Deformation in Liquid Metal}
In the case of fusion liquid walls, there is no external magnetic field present. The current from the wire flows radially inside the liquid metal, as shown in Figure \ref{fig:1.3}. This is a simplified model focusing exclusively on the effects of the radial current flowing through the liquid metal. Current is injected into the liquid metal by applying potential difference between the inner and outer cylindrical surfaces. The setup for this computational model consists of a lower section filled with electrically conductive liquid metal and an upper section with an electrically non-conductive fluid.

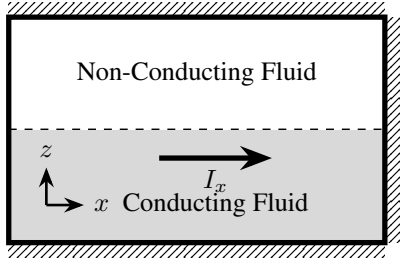
\begin{figure}[H]
	\centering
	\begin{tikzpicture}[scale=1.0, >=Stealth] 
		\def\fluidheight{1.5} 
		
		\coordinate (A) at (0,0);
		\coordinate (B) at (5,0);
		\coordinate (C) at (5,3);
		\coordinate (D) at (0,3);
		
		\draw[dashed, line width=1pt] (0,\fluidheight) -- (5,\fluidheight);
		
		\fill[gray!30] (0,0) rectangle (5,\fluidheight);
		
		\draw[line width=2pt] (A) -- (B) -- (C) -- (D) -- cycle;
		
		\draw[->, line width=2pt] (2,0.75*\fluidheight) -- ++(1.5,0) node[midway,below] {\(I_x\)};

		\draw[->, line width=1pt] (0.5,0.5) -- ++(0.5,0) node[right] {\(x\)};
		\draw[->, line width=1pt] (0.5,0.5) -- ++(0,0.5) node[above] {\(z\)};
		
		\fill[pattern=north east lines, pattern color=black] (0,0) rectangle (5,-0.2); 
		\fill[pattern=north east lines, pattern color=black] (5,0) rectangle (5.2,3); 
		\fill[pattern=north east lines, pattern color=black] (0,3) rectangle (5,3.2); 
		
		\node at (2.75,\fluidheight/3) {Conducting Fluid};
		\node at (2.5,2.25) {Non-Conducting Fluid};
		
	\end{tikzpicture}
	\caption{Sketch of the computational domain used for the numerical simulation of fusion liquid wall model. The picture is a cross section of an annular cylinder. The left and right surfaces correspond to the inner and outer radii of the cylinder. Boundary labels and the representative computational mesh is presented in  Figure \ref{fig:2}.}
	\label{fig:5.12}
\end{figure}

The shape of the container is described by its height \( H \), inner radius \( r_i \), and outer radius \( r_o \). When electric potentials are applied, an electric current flows through the conductive liquid metal as illustrated in Figure \ref{fig:5.12}. The radial current creates a magnetic field that decays in both the radial and axial directions. This spatially varying magnetic field generates a Lorentz force, causing the interface between the two fluids to deform. The expression for the current density associated with radial current has been previously derived and is given by the following equation:

\begin{equation*}
	J(r) = - \frac{\sigma V_o}{r \ln (r_i/r_o)}
\end{equation*}

The total current passing through the liquid metal is represented by $ I = J(r) \times A(r) $. Here, \( A(r) \) denotes the cross-sectional area of the cylinder at a given radius within the liquid metal container. For a liquid metal with a depth of $d$, the total current can be calculated using the expression for current density.

\clearpage

\begin{equation*}
	I =  - \left(\frac{\sigma V_o}{r \ln \left(\frac{r_i}{r_o}\right)}\right)*\left(2 \pi r d \right)
\end{equation*}

Given the total current \( I \), the expression for \( V_o \) is derived from the above equation.

\subsection{Governing equations}
The continuity and momentum equation can be expressed as follows:
\begin{equation*}
	 \nabla \cdot \mathbf{U}  = 0
\end{equation*}

\begin{equation*}
	\frac{\partial \rho \mathbf{U}}{\partial t} + \nabla \cdot \left( \rho \mathbf{U} \mathbf{U} \right) - \nabla \cdot \left( \mu \nabla \mathbf{U} \right) + \nabla p  - \rho \textbf{g} - \mathbf{J}\times \mathbf{B} = 0
\end{equation*}

The density of the mixture \( \rho \) is calculated as: $ \rho = \alpha \rho_l + (1-\alpha) \rho_g  $. Here, $\rho_l$ is the density of liquid and $\rho_g$ is the density of gas. Similarly, the mixture's fluid viscosity and electrical conductivity can be calculated.

\begin{equation*}
\mu = \alpha \mu_l + (1-\alpha) \mu_g \hspace{10em} \sigma = \alpha \sigma_l + (1-\alpha) \sigma_g 
\end{equation*}

The equation governing the temporal evolution of the liquid volume fraction is \cite{moukalled2003pressure,weller1998tensorial,deshpande2012evaluating}  :
\begin{equation*}
\frac{\partial \alpha}{\partial t} + 
 \nabla \cdot \left( \alpha \mathbf{U} \right) +
 \nabla \cdot \left( \mathbf{U}_r \alpha (1-\alpha) \right) = 0
\end{equation*}

The artificial compression velocity $\mathbf{U}_r$  is used to maintain a sharp interface\cite{blishchik2021extensive} and can be calculated as follows:

\begin{equation*}
 \mathbf{U}_r = \mathbf{n}_f \text{ $min$ } \left[ C_a \frac{|\Psi|}{|S_f|}, \text{$max$} \left( \frac{|\Psi|}{|S_f|} \right) \right]
\end{equation*}

Here, $\mathbf{n}_f$ represents the normal vector to the cell face, $\Psi$ denotes the mass flux through the cell face, $S_f$ is the area of the cell face, and $C_a$ is the coefficient used to control the thickness of the interface.

For the applied electrical current, the governing equation is:

\begin{equation*}
	\nabla \cdot \sigma \nabla \Phi - \nabla \cdot \left( \sigma \textbf{U} \times \textbf{B}\right) = 0
\end{equation*}

Subsequently, the current density \( \mathbf{J} \) is calculated as:

\begin{equation*}
	\textbf{J} = - \sigma \nabla \Phi  + \sigma \textbf{U} \times \textbf{B}
\end{equation*}
The governing equation for magnetic vector potential is:

\begin{equation*}
	\nabla \cdot \nabla \textbf{A} - \mu_0  \sigma \nabla \Phi  + \mu_0  \sigma  \textbf{U} \times \textbf{B} = 0
\end{equation*}

Using the definition of magnetic vector potential, the magnetic field can be calculated as $\textbf{B} = \nabla \times \textbf{A}$. The governing equations are presented in non-dimensional form as follows :

\begin{equation*}
	\nabla \cdot \sigma \nabla \Phi  -
	Re_m \, \nabla \cdot \left( \sigma \textbf{U} \times \textbf{B}\right)
	= 0
\end{equation*}

\begin{equation*}
	\nabla \cdot \nabla \textbf{A} - 
	\sigma \nabla \Phi +
	Re_m \, \sigma    \left( \textbf{U} \times \textbf{B}\right)	
	= 0 
\end{equation*}

\begin{equation*}
 \nabla \cdot \textbf{U}   = 0
\end{equation*}

\begin{equation*}
	\frac{\partial \rho \textbf{U}}{\partial t} +
	\nabla \cdot \left(\rho \textbf{U} \textbf{U} \right) - 
	\frac{1}{Re} \nabla \cdot \left(  \nabla \textbf{U} \right)+ 
	\nabla p  - 
	\frac{1}{ Fr^2 } \rho  \mathbf{g} -
	\frac{Ha^2}{Re \, Re_m } \, \mathbf{J} \times \mathbf{B} = 0
\end{equation*}

 Here, $Fr = \displaystyle \frac{U_o}{\sqrt{g_o L_o}}$,  $ \displaystyle Ha = B_o L_o\sqrt{\frac{\sigma_o}{\mu}} $, $ Re_m =   \mu_0 \sigma_o U_o L_o$ and 	$ \displaystyle Re = \frac{U_o \rho_o L_o }{ \mu }$. 
 
 The simulation results, obtained by solving the governing equations in non-dimensional form under the low magnetic Reynolds number approximation, correspond to $Re = 1$ and $Fr = 100$.

\begin{table}[H]
    \centering
    \caption{Parameters used for numerical simulation of fusion liquid wall model.}
    \begin{tabular}{lll}
        \toprule
        Parameter & Description & Value  \\
        \midrule
		$r_i$ & Inner radius of cylinder & $2.35\,mm$\\
		$r_o$ & Outer radius of cylinder & $94\,mm$\\
		$L$ & Height of cylinder & $94\,mm$\\
		$d$ & Depth of liquid metal & $47\,mm$\\
		$\sigma_l$ & Electrical conductivity of liquid metal & $1.3 \times 10^{6}  \,S/m \,$ \\
		$\mu_o$ & Magnetic permeability & $4 \pi \times 10^{-7}  \,H/m \,$  \\
		$\mathbf{A}_{t}$ &  Vector potential at top boundary & $0.0\,T\,m$ \\
		$\rho_g$ & Density of air & $1.0 \,\, kg/m^3$  \\
		$\rho_l$ & Density of liquid metal & $8767 \,\, kg/m^3$  \\
		$p$ & Pressure & $1.0 \,\, Pa $\\
        \bottomrule
    \end{tabular}
	\label{tab:5.6}
\end{table}

\begin{table}[H]
    \centering
	\caption{Boundary conditions used for numerical simulation of fusion liquid wall model.}
    \begin{tabular}{lll}
        \toprule
        & \multicolumn{2}{c}{Boundary Condition}\\
        \cmidrule(r){2-3}
        Boundary & Velocity ($ \mathbf{U}$) & Pressure  ($p$) \\
        \midrule
    	Top boundary & Fixed value & Zero normal gradient \\
    	Bottom boundary & Fixed value & Zero normal gradient \\
    	Left boundary  & Zero normal gradient & Zero normal gradient \\
    	Right boundary & Zero normal gradient & Zero normal gradient \\
    	Front boundary & Fixed value & Zero normal gradient \\
    	Back boundary & Fixed value & Zero normal gradient \\
        \bottomrule
    \end{tabular}
	\label{tab:5.7}
\end{table}
\begin{table}[H]
    \centering
	\caption{Boundary conditions used for numerical simulation of fusion liquid wall model.}
    \begin{tabular}{lll}
        \toprule
        & \multicolumn{2}{c}{Boundary Condition}\\
        \cmidrule(r){2-3}
        Boundary & Electric Scalar Potential ($\Phi$) & Magnetic Vector Potential ($ \mathbf{A}$) \\
        \midrule
    	Top boundary & Zero normal gradient & Fixed value \\
    	Bottom boundary & Zero normal gradient & Zero normal gradient \\
    	Left boundary  & Zero normal gradient & Zero normal gradient \\
    	Right boundary & Zero normal gradient & Zero normal gradient \\
    	Front boundary & Fixed value & Zero normal gradient \\
    	Back boundary & Fixed value & Zero normal gradient \\
        \bottomrule
    \end{tabular}
	\label{tab:5.8}
\end{table}

In this specific problem, there are no boundary conditions that prescribe velocity; instead, the fluid motion is entirely driven by the Lorentz force. The Lorentz force is balanced by inertial forces as follows:

\[
\nabla \cdot \left( \rho \mathbf{U} \mathbf{U} \right) = \mathbf{J}\times \mathbf{B}
\]

Applying the scale factors for each variable and using Ampere's Law results in: $ \displaystyle \left( \frac{1}{L_o} \right)  \rho_o  U_o^2 = J_o B_o =  J_o (\mu_o  L_o J_o) $

\clearpage

\begin{figure}[H]
	\centering
	\begin{subfigure}[b]{0.49\textwidth}
		\centering
		\includegraphics[width=\textwidth,trim={0.3cm 0.5cm 0.5cm 0.5cm},clip]{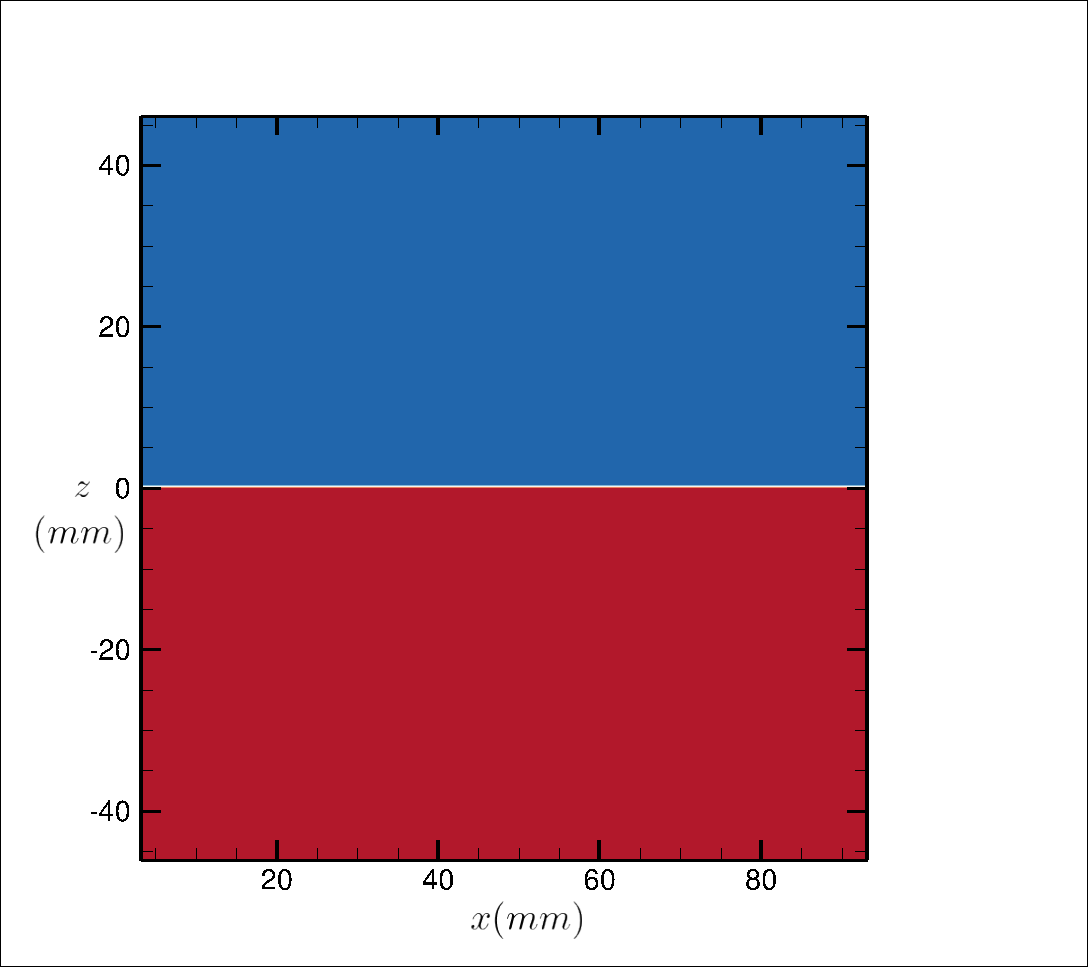}
		\caption{t = 0 }
	\end{subfigure}
	\hfill
	\begin{subfigure}[b]{0.49\textwidth}
		\centering
		\includegraphics[width=\textwidth,trim={0.3cm 0.5cm 0.5cm 0.5cm},clip]{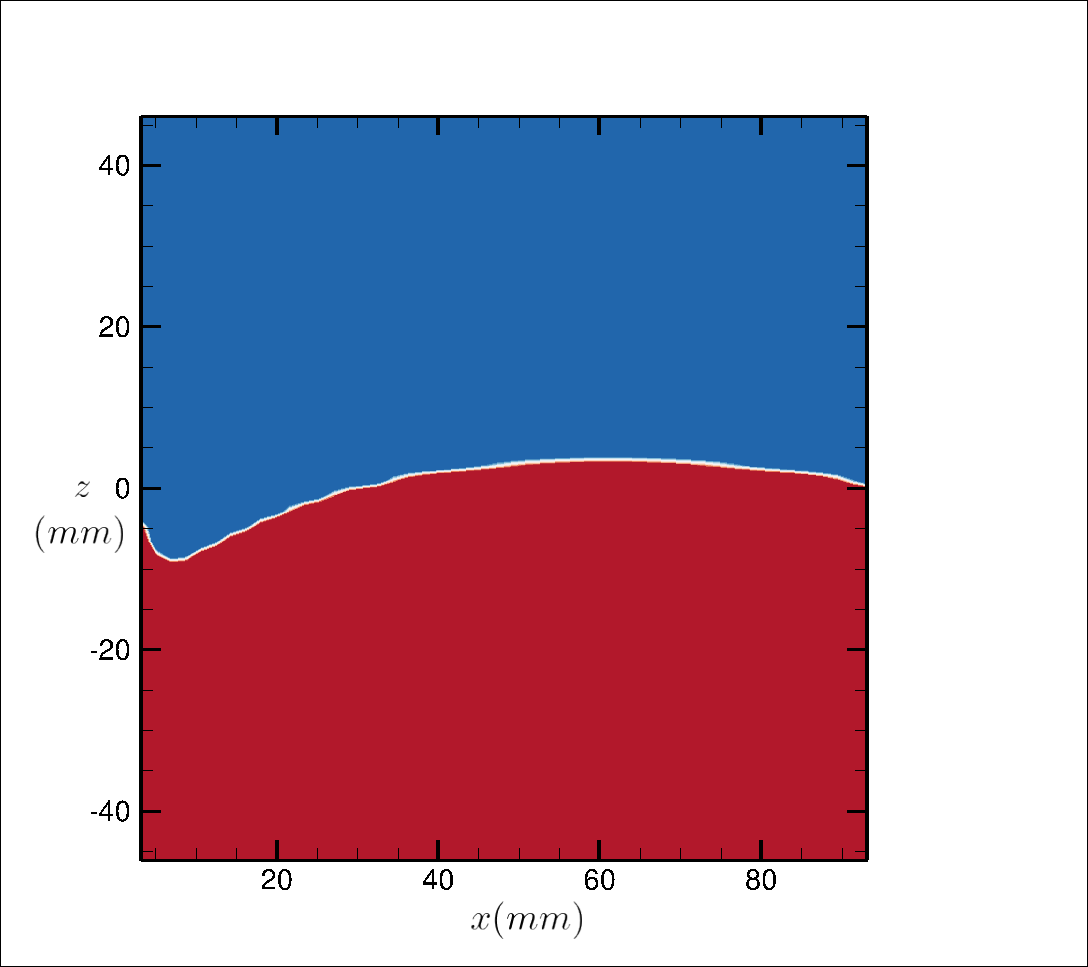}
		\caption{t = 50 }
	\end{subfigure}

	\hfill
	\begin{subfigure}[b]{0.49\textwidth}
		\centering
		\includegraphics[width=\textwidth,trim={0.3cm 0.5cm 0.5cm 0.5cm},clip]{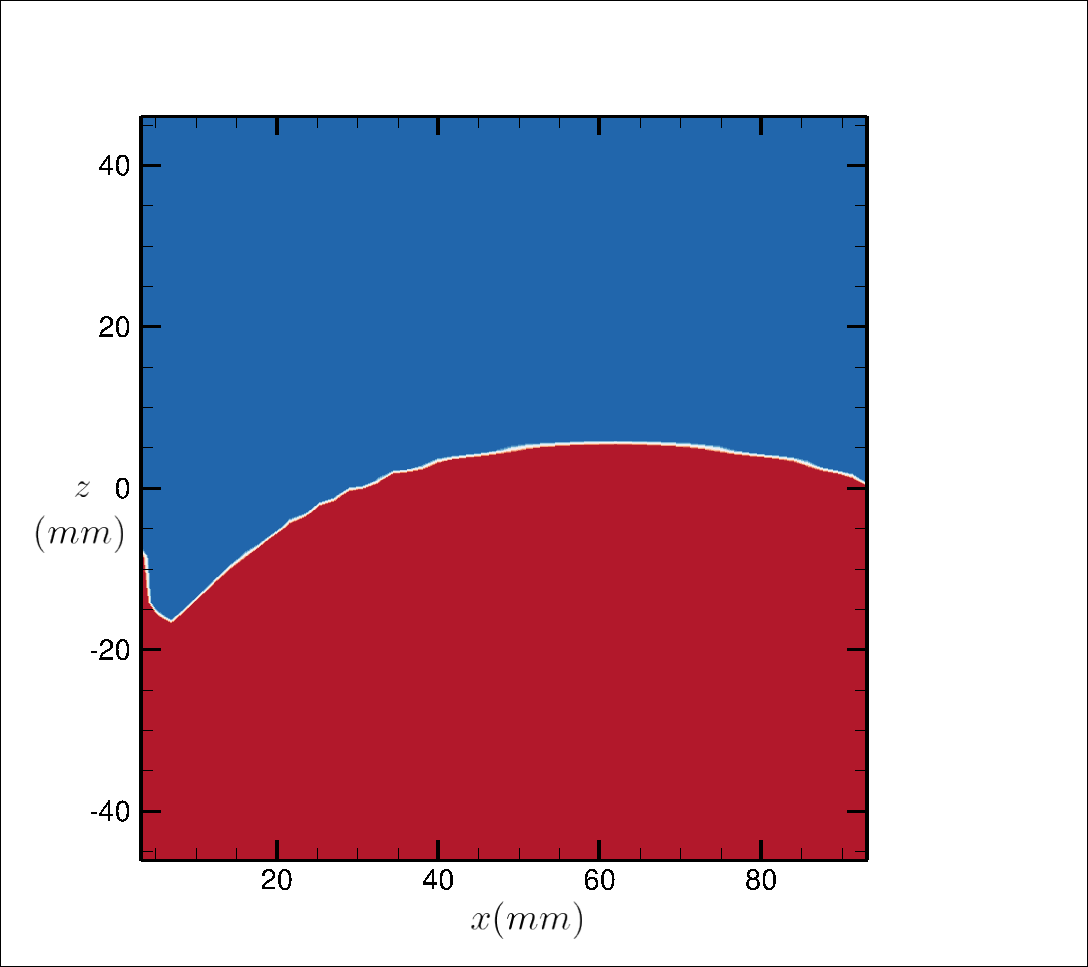}
		\caption{t = 100 }
	\end{subfigure}
	\hfill
	\begin{subfigure}[b]{0.49\textwidth}
		\centering
		\includegraphics[width=\textwidth,trim={0.3cm 0.5cm 0.5cm 0.5cm},clip]{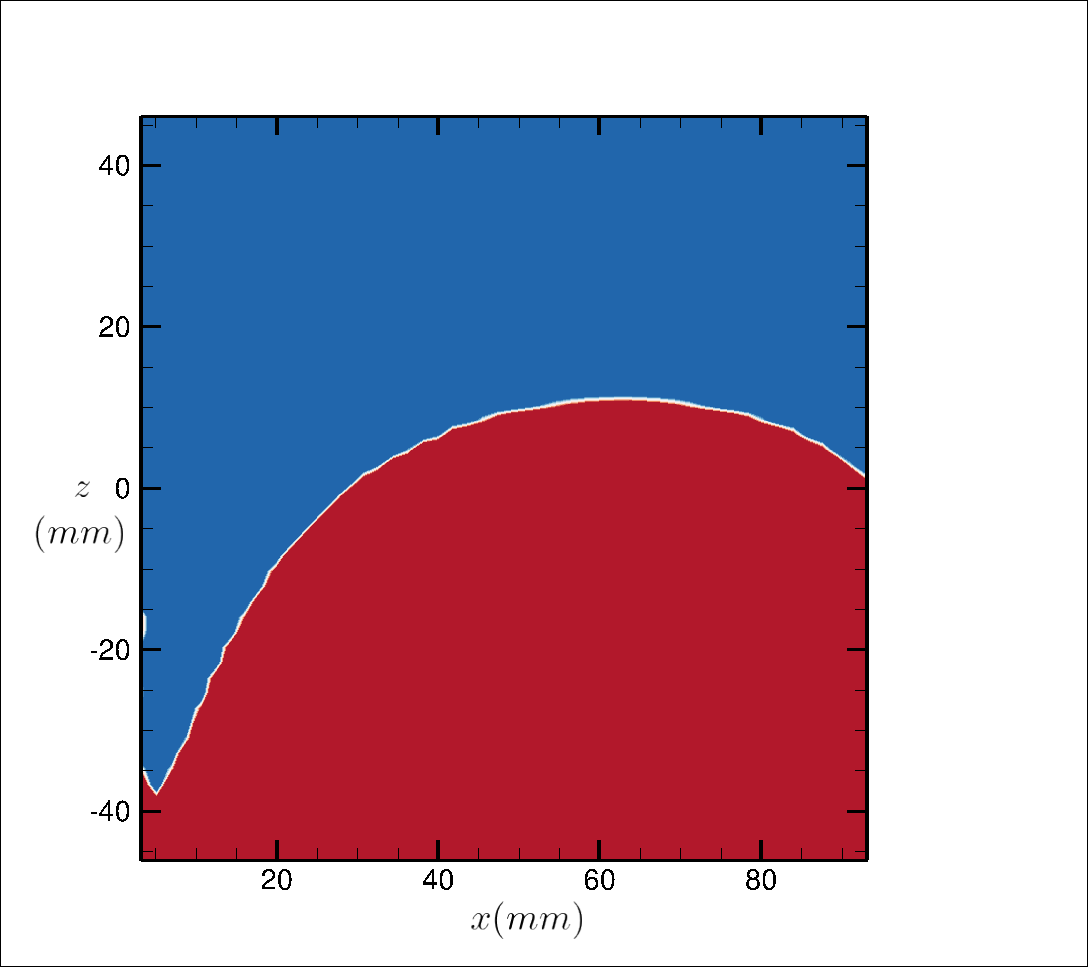}
		\caption{t = 500 }
	\end{subfigure}
	\hfill
	\caption{The figures display evolution of free surface at different non-dimensional times. The red color corresponds to electrically conducting liquid metal corresponds to volume fraction ($\alpha = 1$) and blue corresponds to electrically non-conducting fluid air corresponds to volume fraction ($\alpha = 0$). The initial horizontal free surface at time $t = 0$, deforms due to applied electric current. }
     \label{fig:5.15}
\end{figure}

The velocity scale is given by $ \displaystyle U_o = J_o L_o  \sqrt{ \frac{\mu_o}{ \rho_o} } $, where $J_o$ is current density scale, $L_o$ is length scale, $\mu_o$ is magnetic permeability and $\rho_o$ is density scale. The current density scale can be expressed in-terms of current $I_o$ and characteristic area $A$
\[
U_o = \frac{I_o L_o}{A}   \sqrt{ \frac{\mu_o}{ \rho_o} }
\]
The characteristic area $A$ is taken as $A = 2 \pi r_o L_o$. Here $r_o$ is outer radius of cylinder. The Reynolds number can then be calculated based on this velocity scale:
\[ Re = \frac{U_o \, \rho_o \, L_o}{\mu_{vo}} = \frac{I_o  \, L_o^2}{ A \, \mu_{vo}} \sqrt{ \mu_o  \, \rho_o } \]
This Reynolds number definition is convenient as it correlates directly with the applied current. Here \( \mu_{vo} \) is the viscosity scale factor for \( \mu_v \). Using these scale factors, the non-dimensional group $N$ becomes 1, since magnetic field is generated by applied current itself and there are no external magnetic fields. 

\begin{equation*}
	N = \frac{Ha^2  }{ Re \, Re_m} = 1
\end{equation*}

\begin{figure}[H]
	\centering
	\begin{subfigure}[b]{0.49\textwidth}
		\centering
		\includegraphics[width=\textwidth,trim={0.3cm 0.5cm 0.5cm 0.5cm},clip]{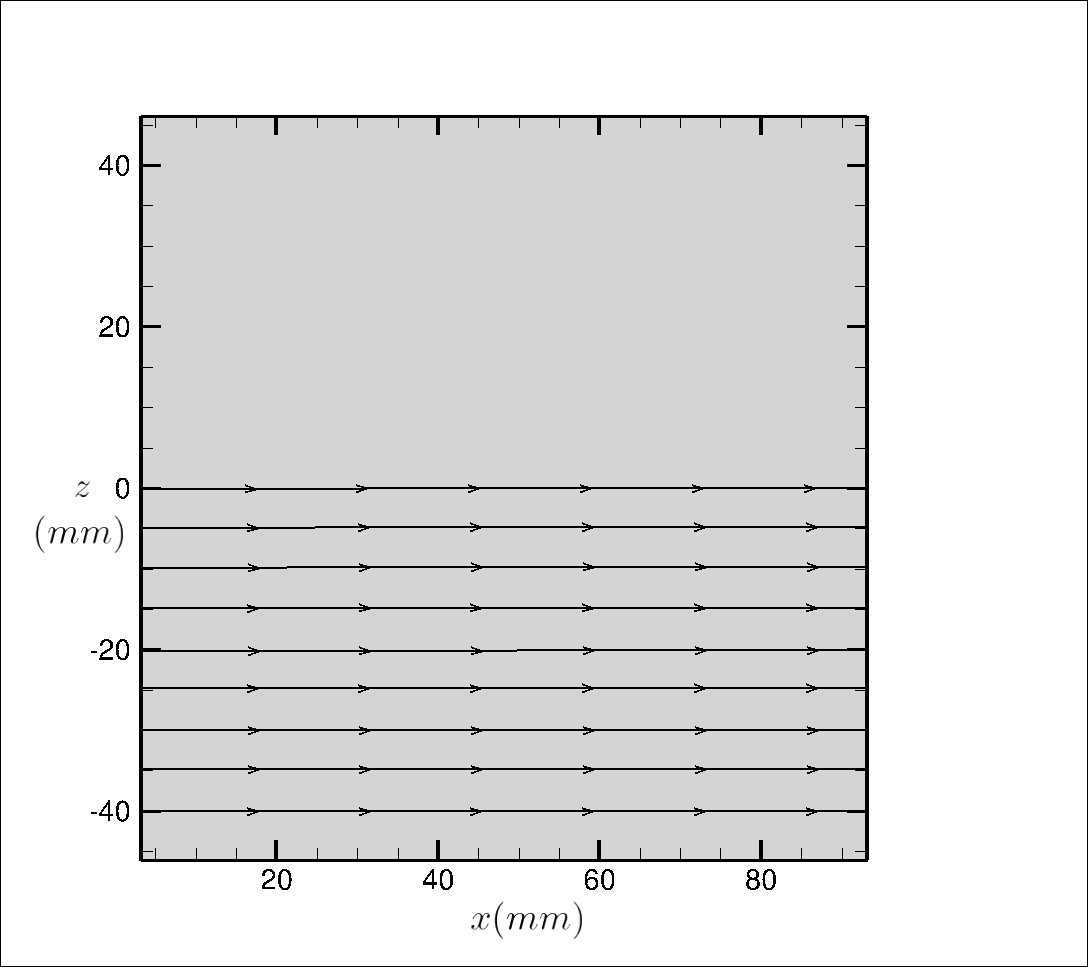}
		\caption{t = 1 }
	\end{subfigure}
	\hfill
	\begin{subfigure}[b]{0.49\textwidth}
		\centering
		\includegraphics[width=\textwidth,trim={0.3cm 0.5cm 0.5cm 0.5cm},clip]{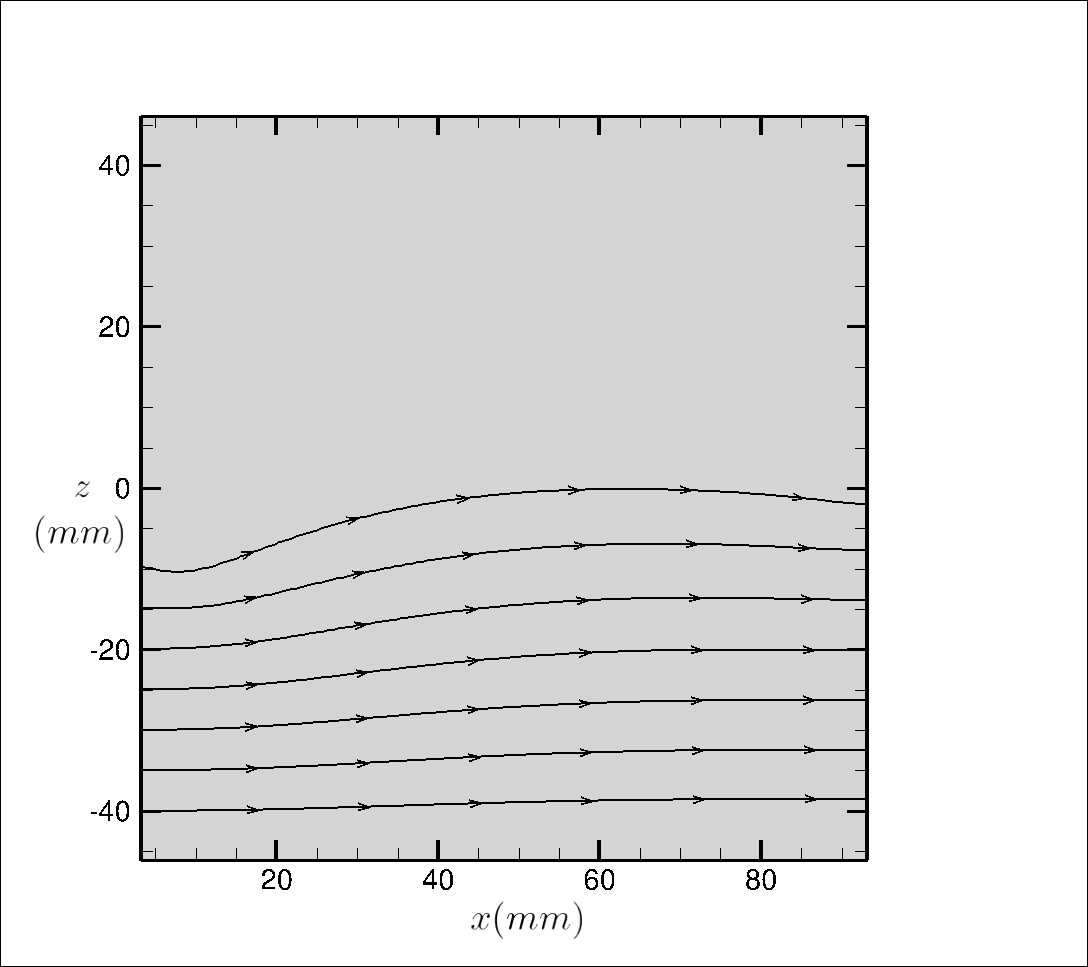}
		\caption{t = 50 }
	\end{subfigure}
	
	\hfill
	\begin{subfigure}[b]{0.49\textwidth}
		\centering
		\includegraphics[width=\textwidth,trim={0.3cm 0.5cm 0.5cm 0.5cm},clip]{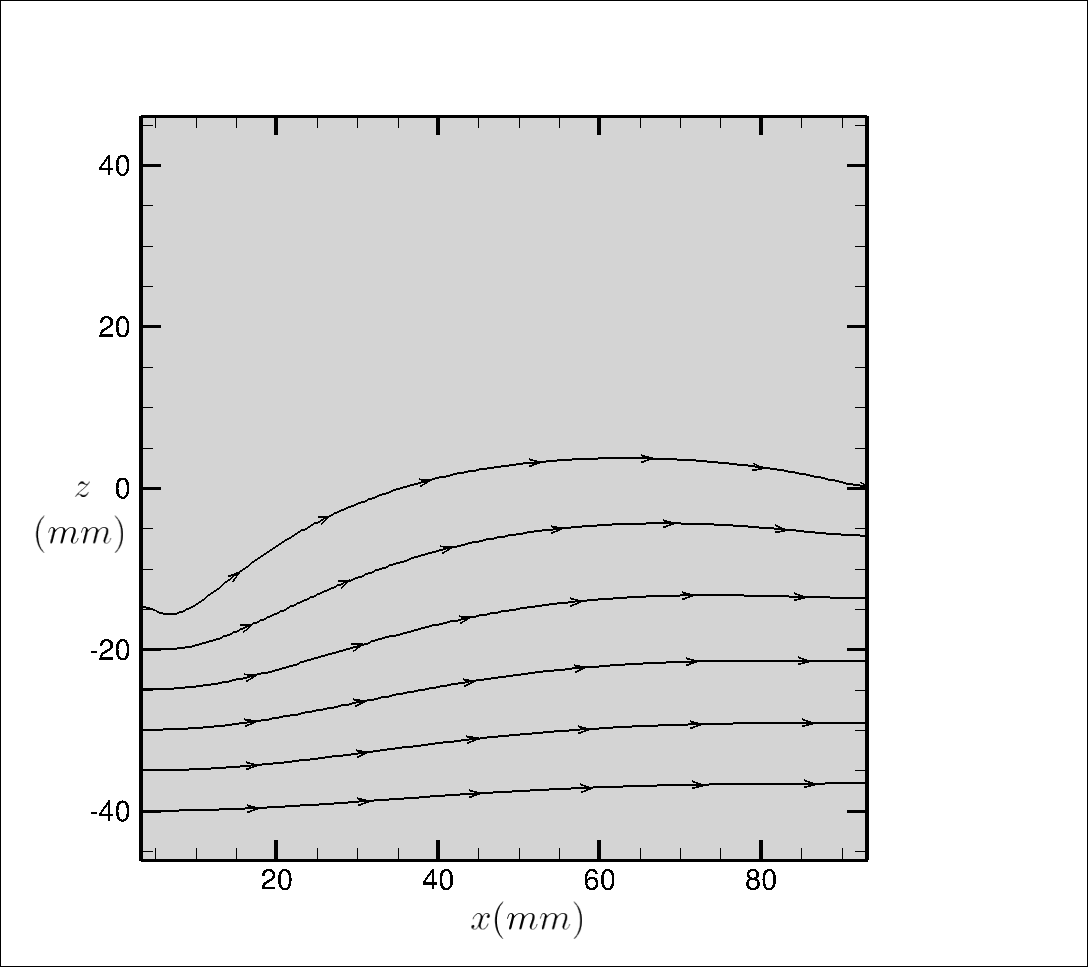}
		\caption{t = 100 }
	\end{subfigure}
	\hfill
	\begin{subfigure}[b]{0.49\textwidth}
		\centering
		\includegraphics[width=\textwidth,trim={0.3cm 0.5cm 0.5cm 0.5cm},clip]{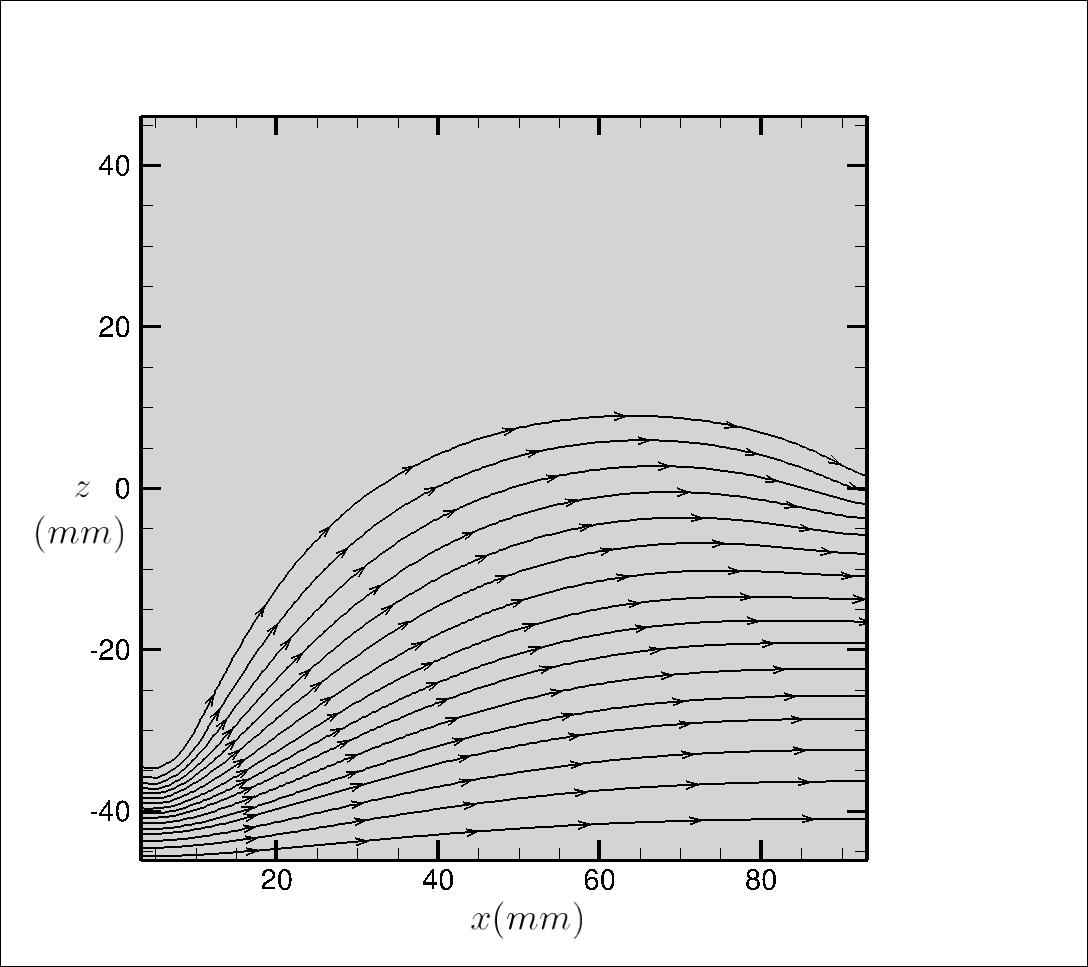}
		\caption{t = 500 }
	\end{subfigure}
	\hfill
	\caption{The figures display evolution of current density streamlines at different non-dimensional times. }
 	\label{fig:5.17}
\end{figure}

\subsection{Mechanism of Free surface deformation}

The mechanism driving the deformation of the free surface can be described as follows: 

\begin{enumerate}
    \item An electric potential difference is applied across the cylinder's inner and outer cylindrical surfaces. Current ($\mathbf{J} $) flows radially within the liquid metal due to applied potential difference as shown in Figure \ref{fig:5.17}. 
    \item Radial current flowing through the liquid metal generates azimuthal magnetic field ($\mathbf{B}$) in accordance with Ampere's law. The magnetic field due to radial current is presented in Figure \ref{fig:5.21}.
    \begin{figure}[H]
	\centering
	\begin{subfigure}[b]{0.49\textwidth}
		\centering
		\includegraphics[width=\textwidth,trim={0.3cm 0.5cm 0.5cm 0.5cm},clip]{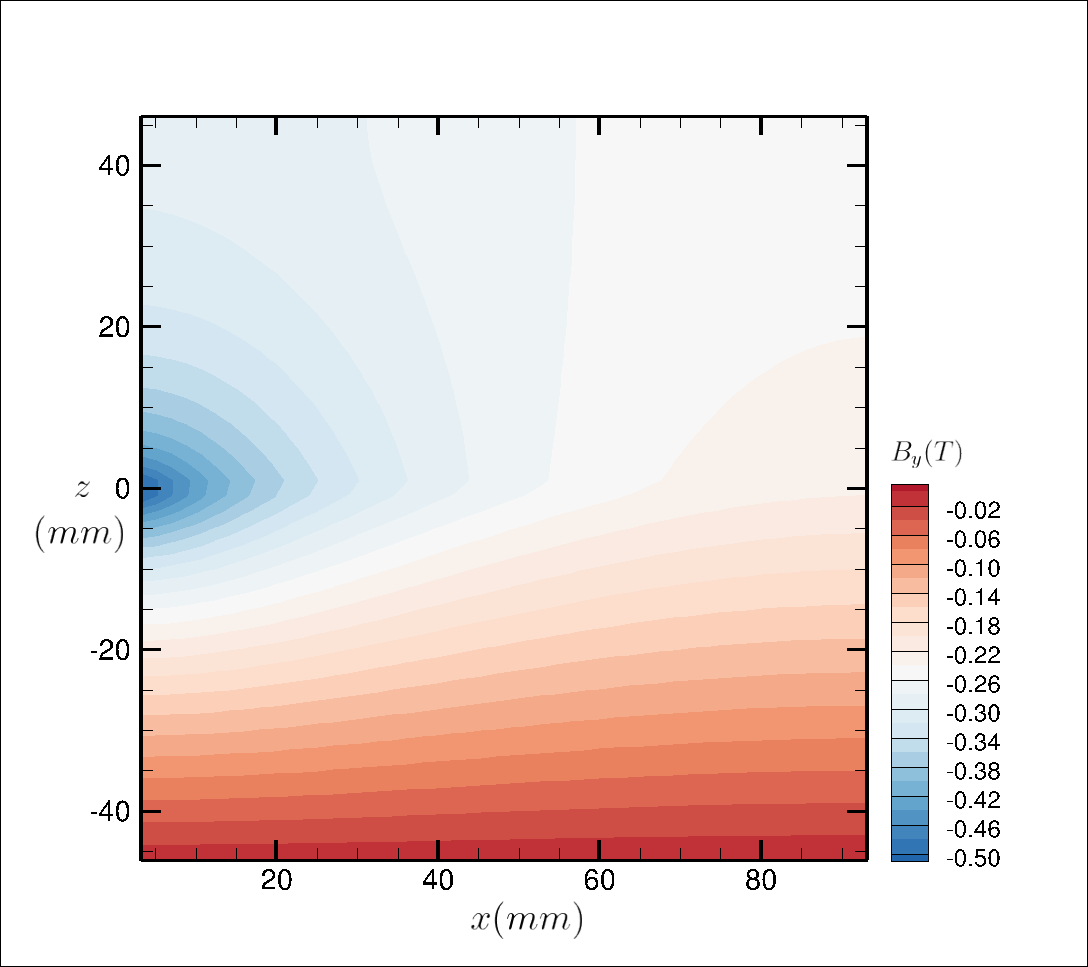}
		\caption{t = 1 }
	\end{subfigure}
	\hfill
	\begin{subfigure}[b]{0.49\textwidth}
		\centering
		\includegraphics[width=\textwidth,trim={0.3cm 0.5cm 0.5cm 0.5cm},clip]{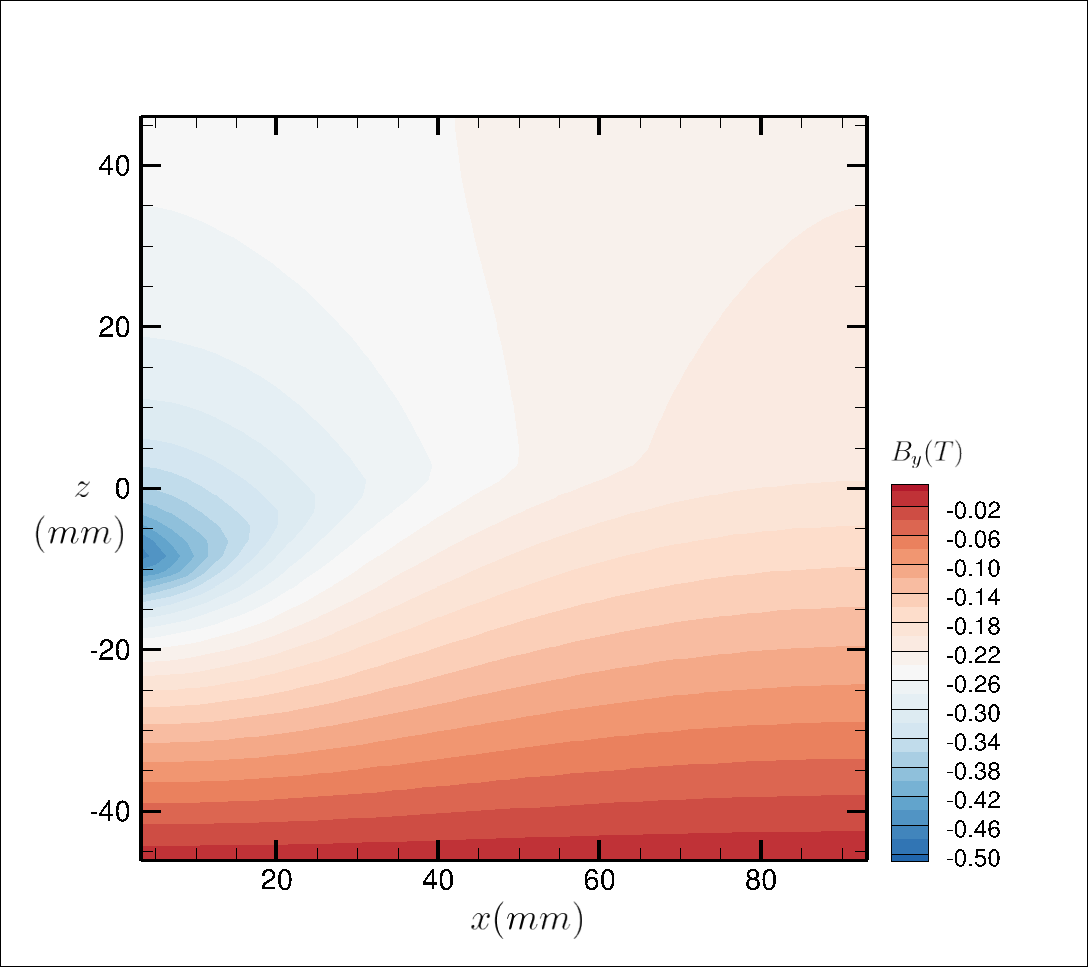}
		\caption{t = 50 }
	\end{subfigure}
	
	\hfill
	\begin{subfigure}[b]{0.49\textwidth}
		\centering
		\includegraphics[width=\textwidth,trim={0.3cm 0.5cm 0.5cm 0.5cm},clip]{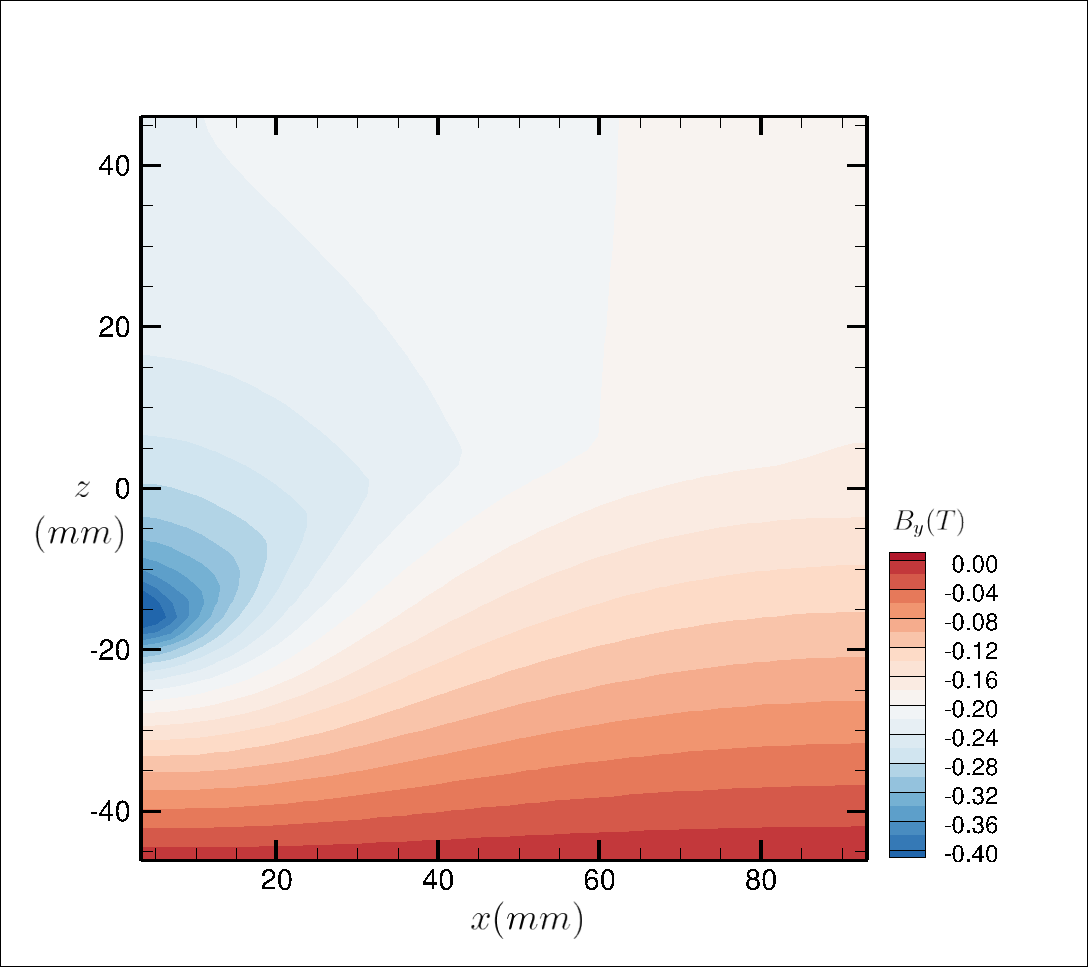}
		\caption{t = 100 }
	\end{subfigure}
	\hfill
	\begin{subfigure}[b]{0.49\textwidth}
		\centering
		\includegraphics[width=\textwidth,trim={0.3cm 0.5cm 0.5cm 0.5cm},clip]{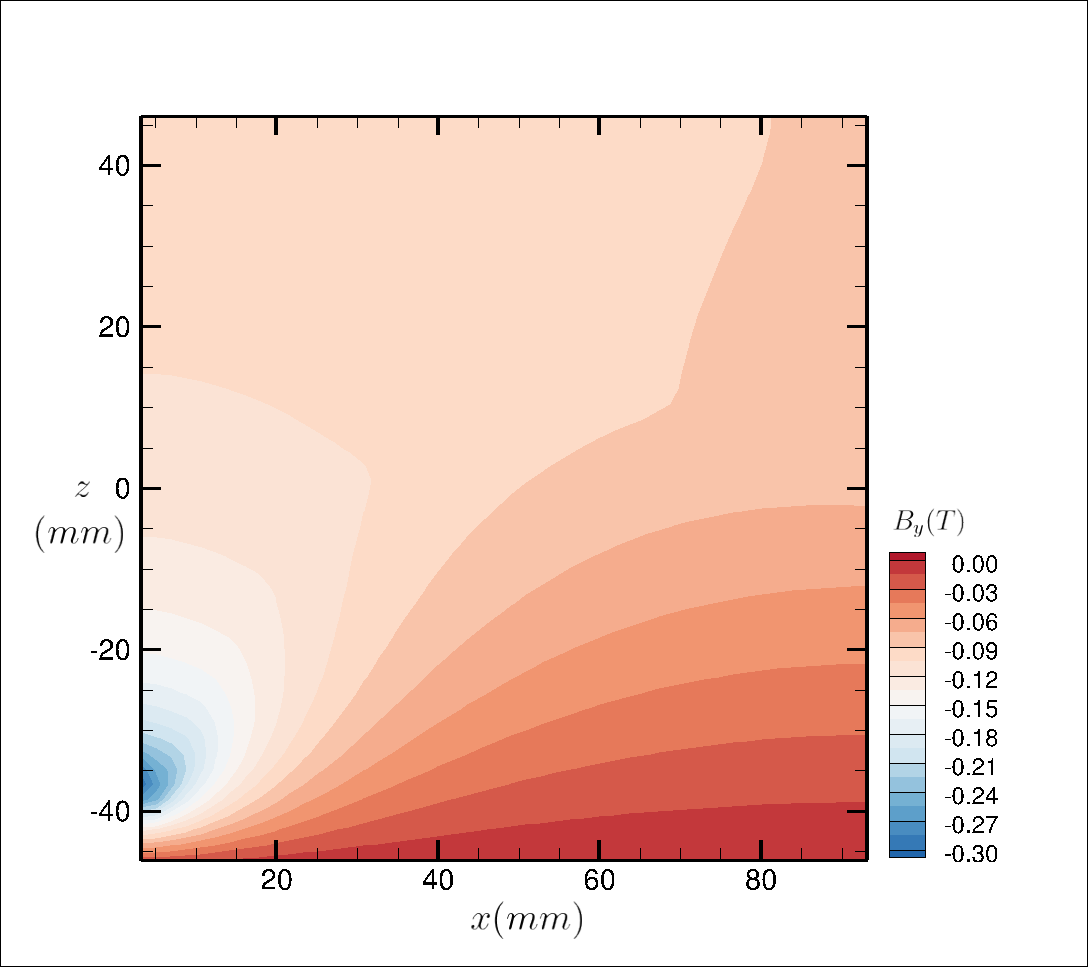}
		\caption{t = 500 }
	\end{subfigure}
	\hfill
	\caption{The figures display evolution of $y (\theta) $ component of magnetic field at different non-dimensional times. }
     \label{fig:5.21}
\end{figure}
    \item The radial current, in conjunction with the azimuthal magnetic field, produces a Lorentz force $\mathbf{J} \times \mathbf{B}$, whose direction is downward as determined by the cross product $\mathbf{J} \times \mathbf{B}$.
    \item The cross-sectional area through which current flows increases radially, given by $2 \pi r d$. Here, $r$ represents the radial distance, and $d$ denotes the depth of the liquid metal. As the radial distance $r$ increases, the cross-sectional area also increases. Consequently, the current density must decay in the radial direction to satisfy the continuity equation $\nabla \cdot \mathbf{J} = 0$.
    \item As the current density decreases radially, the magnetic field also decreases radially. Additionally, the magnetic field also decreases axially as the enclosed current decreases in the axial direction \cite{fleisch2008student,griffiths2017introduction,jackson1999classical}, starting from the free surface . The enclosed current can be conceptualized as the number of current density lines crossing a hollow cylindrical surface at a specific radial distance with different heights. A cylinder with its base coinciding with the bottom surface of the container will have fewer current lines crossing it if its height is much smaller than the depth of the liquid metal. Conversely, if the height of the cylinder approaches the depth of the metal, and its top surface coincides with the free surface, the number of enclosed current density lines is maximum. Consequently, in the axial direction, the magnetic field is stronger near the free surface, whereas in the radial direction, it is stronger near the inner cylindrical surface. Thus, a strong magnetic field is observed at the intersection of the interface with the inner cylindrical surface. This two dimensional magnetic field can be visualized in Figure \ref{fig:5.21}.
    \item Since both the current density and magnetic field decay in the radial direction, the resulting net downward Lorentz force $\mathbf{J} \times \mathbf{B}$ is stronger near the inner cylindrical surfaces and decays radially. Consequently, the liquid metal is pushed downward near the inner cylindrical surface with a magnitude stronger than at other radial positions. Consequently, fluid pressure builds up in response to this body force, leading to deformation of the liquid metal's free surface as shown in Figure \ref{fig:5.15}.
    \item As the liquid metal deforms, the current density lines and resulting magnetic field adjust to the shape of the liquid metal. The simulation is stopped before the liquid metal loses contact with the inner cylindrical surface.
\end{enumerate}

\section{Summary and Conclusion}
An attempt has been made to numerically simulate the free surface deformation of a fusion liquid wall, caused by the injected current from Z-pinch plasma. The approach is not only applicable to fusion wall scenarios but broadly to situations involving current injection into any conducting liquid. The development of the final solver involved a systematic approach, beginning with solving Maxwell's equations, then integrating them into the single-phase Maxwell-Navier-Stokes framework, and finally extending to the two-phase Maxwell-Navier-Stokes equations. The potential formulation of Maxwell's equations is solved using a finite volume numerical method and then compared with the derived analytical solution for an axially current-carrying wire. The numerical solutions are found to be in close agreement with the analytical solutions. The study then extends to radial current flow in a cylindrical domain, where the current flows radially in an annular cylinder from the inner cylindrical surface to the outer cylindrical surface. This simulation is particularly relevant as it closely relates to the current flow in the fusion liquid wall's case, aiding in understanding the boundary conditions for simulating electromagnetic fields in such scenarios. In the case of an axial current-carrying wire, the current density is uniform. However, for radial current flow, the current density decays radially. Additionally, while the magnetic field in the axial current-carrying wire varies only radially, in the radial current case, it changes both radially and axially.  

In scenarios where liquid metals flow in the presence of electromagnetic fields, secondary electrical currents and magnetic fields are induced. To accommodate this, Maxwell's equations in their potential form are extended to moving conductors, utilizing Ohm's law for such moving conductors. Maxwell's equations for moving conductors are then solved in conjunction with the Navier-Stokes equations, which include a Lorentz body force term, arising due to electromagnetic fields. The combined Maxwell-Navier-Stokes equations are formulated for two specific scenarios: one involving purely applied current, and the other a combination of applied current and an external magnetic field. To validate the Maxwell-Navier-Stokes system for single-phase flows, electrically driven flows in an annular cylinder is simulated. This test case references Khalzov's dissertation \cite{khalzov2008equilibrium}, where the evolution of the magnetic field is solved using toroidal field components and poloidal flux functions. In contrast, this study solves the same problem using the potential formulation of Maxwell's equations in Cartesian coordinates. The setup of the case is as follows: radial currents flow in an annular cylinder from the inner cylindrical surface to the outer cylindrical surface. Additionally, an axial magnetic field permeates the entire domain. The interaction between the radial currents and the axial magnetic field generates an azimuthal Lorentz force, causing the metal to spin. This phenomenon is simulated, and the resulting angular momentum profile shows good agreement with the analytical solution derived from Khalzov's work \cite{khalzov2008equilibrium}.

The final study focuses on simulating the two-phase Maxwell-Navier-Stokes equations with the goal of including free surface deformation, representing a fusion liquid wall model scenario. In this model, current from a wire, analogous to the plasma column in a Z-pinch, flows radially through the liquid metal layer.  However, the wire or plasma column itself is not part of the model; only the radial current flowing into the liquid metal layer is considered. Unlike the previous study on electrically driven flow in an annular channel, there is no external magnetic field present to rotate the flow. In this scenario, the radial current flow results in a non-uniform current density along the radius, decaying in the radial direction. Consequently, the magnetic field generated by this current also decays radially. Furthermore, there is an axial variation in the magnetic field, as the enclosed current changes axially, leading to a two-dimensional variation in the magnetic field. This spatially varying magnetic field produces a net downward force within the liquid metal. The resulting Lorentz force leads to a buildup of fluid pressure, causing deformation of the interface, a phenomenon demonstrated through numerical simulation.

\bibliographystyle{unsrt}  
\bibliography{references}

\end{document}